\documentclass[12pt]{article}
\usepackage{a4wide}
\usepackage[normalem]{ulem}
\usepackage[dvipdfmx]{graphicx}
\usepackage[top=3.5cm,bottom=3.5cm,left=2cm,right=2cm]{geometry}
\usepackage{amsmath,amssymb}

\usepackage{color}

\usepackage{slashed}
\usepackage[bookmarks=true,bookmarksnumbered=true,setpagesize=false]{hyperref}

\usepackage{comment}

\newcommand{\be}{\begin{equation}}
\newcommand{\e}{\end{equation}}
\newcommand{\aln}[1]{\begin{align}#1\end{align}}

\newcommand{\ov}{\over}

\begin{document}
\title{
\vspace{-2cm}
\vbox{
\baselineskip 14pt
\hfill \hbox{\normalsize OU-HET-1130
}} 
\bf\Large 
Gravitational waves\\ in models with multicritical-point principle
}

\author{
Yuta Hamada,\thanks{E-mail: \tt yhamada@fas.harvard.edu}\,
Hikaru~Kawai,\thanks{E-mail: \tt hikarukawai@phys.ntu.edu.tw}\, 
Kiyoharu Kawana,\thanks{E-mail: \tt kawana@snu.ac.kr}\,
Kin-ya Oda,\thanks{E-mail: \tt odakin@lab.twcu.ac.jp}\, and
Kei Yagyu\thanks{E-mail: \tt yagyu@het.phys.sci.osaka-u.ac.jp}
\bigskip\\
\normalsize
\it 
 $^*$  Department of Physics, Harvard University, Cambridge, MA 02138 USA,\\
\normalsize
 \it
 $^{\dagger}$ Department of Physics and Center for Theoretical Physics,
 \\
 \normalsize
\it 
  National Taiwan University, Taipei, Taiwan 106,
  \\
 \normalsize 
 \it 
  Physics Division, National Center for Theoretical Sciences, Taipei 10617, Taiwan 
 \\ 
\normalsize 
\it  
$^{\ddag}$  Center for Theoretical Physics, Department of Physics and Astronomy,\\
 \normalsize
\it  Seoul National University, Seoul 08826, Korea,
\\
\normalsize
\it  $^{\S}$ Department of Mathematics, Tokyo Woman's Christian University, Tokyo 167-8585, Japan
\\
\normalsize
\it  $^{\P}$ Department of Physics, Osaka University, Toyonaka, Osaka 560-0043, Japan
}


\maketitle
\vspace{-1cm}
\begin{abstract}\noindent
The multicritical-point principle (MPP) provides a natural explanation of the large hierarchy between the Planck and electroweak scales. We consider a scenario in which MPP is applied to the Standard Model extended by two real singlet scalar fields $\phi$ and $S$, and a dimensional transmutation occurs by the vacuum expectation value of $\phi$. In this paper, we focus on the critical points that possess a $\mathbb Z_2$ symmetry $S\to-S$ and all the other fields are left invariant. Then $S$ becomes a natural dark matter (DM) candidate. Further, we concentrate on the critical points where $\phi$ does not possess further $\mathbb Z_2$ symmetry so that there is no cosmological domain-wall problem. 
Among such critical points, we focus on maximally critical one  called CP-1234 that fix all the superrenormalizable parameters. 
We show that there remains a parameter region that satisfies the DM relic abundance, DM direct-detection bound and the current LHC constraints. 
In this region, we find a first-order phase transition in the early universe around the TeV-scale temperature. The resultant gravitational waves are predicted with a peak amplitude of ${\cal O}(10^{-12})$ at a frequency of $10^{-2}\text{--}10^{-1}$\,Hz, which can be tested with future space-based instruments such as DECIGO and BBO.

\mbox{}
\end{abstract}
%

\section{Introduction}
Gravitational Wave (GW) astronomy~\cite{LIGOScientific:2016aoc,LIGOScientific:2017vwq,LIGOScientific:2018mvr} is one of the most fascinating research fields because it allows us to explore the physics of the early universe, complementary to the cosmic microwave background (CMB) observations. Gravitational waves from compact binaries~\cite{LIGOScientific:2021sio,LIGOScientific:2021jlr,LIGOScientific:2017adf} and stochastic background~\cite{NANOGrav:2020bcs,Blasi:2020mfx,Vaskonen:2020lbd,DeLuca:2020agl,Nakai:2020oit,Addazi:2020zcj} allow us to test new physics models as well as gravity theories.  
In particular, stochastic GWs originating from first-order phase transitions (FOPTs)~\cite{Grojean:2006bp,Leitao:2012tx,Jinno:2015doa,Hashino:2016rvx} have received much attention in recent years, as many new physics models predict them in the early Universe, contrary to the pure Standard Model (SM), which shows a crossover electroweak (EW) transition~\cite{Kajantie:1996mn,Laine:1998jb}.

In this paper, we study the FOPT and the resultant GW signals in an extension of the SM by two real singlet scalars $\phi$ and $S$ that account for the hierarchy between the EW and Planck scales~\cite{Hamada:2020wjh,Hamada:2021jls}.  
In the original Coleman-Weinberg (CW) mechanism~\cite{Coleman:1973jx,Kawai:2021lam}, the scalar field has a loop-suppressed mass compared to its vacuum expectation value (VEV), and cannot be regarded as the observed Higgs boson at LHC. 
That is why $\phi$ must be introduced in addition to the SM Higgs as a scale-generating scalar.
On the other hand, the role of the gauge field in the original CW model can be played by~$S$~\cite{Haruna:2019zeu}. 
In this sense, this model~\cite{Hamada:2020wjh,Hamada:2021jls} can serve as a minimal model of the scale generation. 
As a bonus, we can naturally identify $S$ as the weakly-interacting-massive-particle dark matter 
by imposing the $\mathbb{Z}_2^{}$ symmetry $S\to-S$, with all other fields being invariant.

The starting assumption of the CW mechanism is the absence of dimensionful parameters, especially the scalar mass-squared, in the tree-level Lagrangian. 
This assumption is sometimes called the classical conformality or the classical scale invariance~\cite{Bardeen:1995kv,Meissner:2006zh,Meissner:2008gj,Foot:2007iy,Iso:2009ss,Iso:2009nw,Hur:2011sv,Iso:2012jn,Englert:2013gz,Hashimoto:2013hta,Holthausen:2013ota,Hashimoto:2014ela,Kubo:2014ova,Endo:2015ifa,Kubo:2015cna,Jung:2019dog}, and can be naturally explained~\cite{Haruna:2019zeu,Kawai:2021lam} as a consequence of the multicritical-point principle (MPP)~\cite{Bennett:1993pj,Froggatt:1995rt}.
There are several ways to describe the MPP, but a simple one is ``the coupling constants of a theory should be tuned to one of the multicritical points at which some of the extrema of the low-energy effective potential are degenerate'';
see Refs.~\cite{Nielsen:2012pu,Kawai:2011qb,Kawai:2013wwa,Hamada:2014ofa,Hamada:2014xra,Hamada:2015dja,Hamada:2020wjh,Kannike:2020qtw,Hamada:2021jls,Kawai:2021lam,Racioppi:2021ynx} for further discussion.
In this sense, the MPP determines the couplings constants in the low-energy effective potential, and in particular, provides a natural answer to the fine-tuning problem.

Some of the mechanisms for realizing the MPP prefer criticalities with larger multiplicity~\cite{Nielsen:2012pu,Kawai:2013wwa,Kawai:2021lam}. 
In this paper, for the first time, we consider the two-scalar extension of the SM with maximal multicriticality in the sense that all the superrenormalizable couplings of the effective potential 
are fixed to a multicritical point, that is, all the relevant parameters in the low energy effective theory are fixed by the MPP. 

As a first attempt, we focus on a class of multicritical points at which the Higgs mass-squared vanishes along with its superrenormalizable couplings to $\phi$ or $S$. 
The problem is then reduced to finding a maximally multicritical point of the two-scalar subsystem. 
Indeed in Ref.~\cite{Kawai:2021lam}, the maximally multicritical points of the two-scalar system with the $\mathbb{Z}_2$ symmetry for $S$ have already been classified. 
Among them, the so-called {\bf 1234} case (CP-1234 in this paper) is of particular interest because it can realize the FOPT without domain-wall problems around the TeV scale.

The organization of the paper is as follows.
In Section~\ref{Model}, we discuss the maximal criticality at CP-1234 for the two-scalar extension of the SM.
In Section~\ref{sec:FOPT}, we study the FOPT by computing the bounce action and by obtaining the critical, nucleation, and percolation temperatures from it.
In Section~\ref{sec:GW}, we study the GW signals from the FOPT.
We find that the phase-transition strength parameters are typically $\alpha=10^{-2}\text{--}10^{-1}$ and $\beta/H=10^{2}\text{--}10^3$. These values imply that the FOPT is not much strong and the resultant GW signals are dominated by the sound wave contributions. 
The resultant peak amplitude can be as large as ${\cal O}(10^{-12})$ around the frequency $10^{-2}\text{--}10^{ -1}\,$Hz, which can be tested by DECIGO and BBO.  
In Section~\ref{summary}, we provide a summary.

\section{Model}\label{Model}
We consider a model with two real singlet scalar fields $\phi$ and $S$ in addition to the SM particles. To make $S$ a dark matter candidate, we impose an unbroken $\mathbb{Z}_2$ symmetry whose transformation property is defined as $S \to -S$ and all the other fields to be even.\footnote{
This $\mathbb Z_2$ invariance can be viewed as a consequence of the MPP, as well as the maximal multicriticality that will be discussed below. In general, the existence of an accidental low-energy global symmetry can be viewed as a realization of multicriticality.
}

The tree-level Lagrangian is
\aln{
{\cal L}={\cal L}_{\rm SM} &+\frac{1}{2}(\partial_\mu \phi)^2+\frac{1}{2}(\partial_\mu S)^2 - V(H, \phi,S), 
\label{lagrangian}
}
where ${\cal L}_{\rm SM} $ is the SM Lagrangian without the Higgs potential for the SM doublet $H$.
The scalar potential is generally given by
\aln{
V(H, \phi,S)&=\frac{\lambda_H}{2} (H^\dagger H)^2
+\frac{\lambda_\phi^{}}{4!}\phi^4 + \frac{\lambda_{\phi S} }{4}\phi^2S^2 + \frac{\lambda_S }{4!}S^4 - \frac{\lambda_{\phi H} }{2}\phi^2(H^\dagger H)
+\frac{\lambda_{SH}}{2}S^2(H^\dagger H) \notag\\
&\quad+\mu_H^2(H^\dagger H) + \frac{\mu_S^2}{2}S^2 + \mu_{\phi H}\phi H^\dagger H + {\mu_{\phi S}\over2}\phi S^2+\mu_1^{3}\phi + \frac{\mu_2^{2}}{2}\phi^2 + \frac{\mu_3^{}}{3!}\phi^3. \label{potential}
}
Here, the masses and couplings are the renormalized ones that appear in the low energy effective potential; we will discuss the concrete renormalization scheme in Eq.~\eqref{one-loop potential} and below.

We assume the \emph{maximal} multicriticality, in which \emph{all} the superrenormalizable parameters, i.e., the $\mu_i$ parameters given in the second line of Eq.~(\ref{potential}), 
are fixed to one of the multicritical points in the parameter space by the MPP.

In particular, we focus on a class of multicritical points in which 
$(\mu_H,\mu_S,\mu_{\phi H},\mu_{\phi S},\mu_1,\mu_2,\mu_3)$ are fixed to be $(0,0,0,0,\mu_1,\mu_2,\mu_3)$.
The vanishing of the first four dimensionful parameters can be understood as follows.
In the field space of $(H,S,\phi)$, we focus on the region where $H$ and $S$ are small, and rewrite the potential~\eqref{potential} as
\aln{
V(H, \phi,S)&=\frac{\lambda_H}{2} (H^\dagger H)^2
+\frac{\lambda_\phi^{}}{4!}\phi^4  + \frac{\lambda_S }{4!}S^4
+\frac{\lambda_{SH}}{2}S^2(H^\dagger H)\nonumber\\
&\quad- \frac{\lambda_{\phi H} }{2}\left[(\phi-\tilde{\mu}_H')^2 + \tilde{\mu}_H^2\right](H^\dagger H)
+ \frac{\lambda_{\phi S} }{4}\left[(\phi - \tilde{\mu}_S')^2 + \tilde{\mu}_S^2\right]S^2\nonumber\\
&\quad  +\mu_1^{3}\phi + \frac{\mu_2^{2}}{2}\phi^2 + \frac{\mu_3^{}}{3!}\phi^3.
}
The point with $\tilde{\mu}_H^2 = \tilde{\mu}_S^2 = 0$ in the parameter space is multicritical in the following sense (see also Ref.~\cite{Hamada:2020wjh}): if $\tilde{\mu}_H^2 <0$, 
then $\left[(\phi-\tilde{\mu}_H')^2 + \tilde{\mu}_H^2\right]$ can change its sign when $\phi$ is varied, and stability and instablity of $H$ around $H=0$ is flipped; on the other hand, if $\tilde{\mu}_H^2>0$, there is no such flip; in this sense $\tilde{\mu}_H^2=0$ is a critical point; the same argument holds for $\tilde{\mu}_S^2$.
Hereafter, we choose such a multicritical point $\tilde{\mu}_H^2 = \tilde{\mu}_S^2 = 0$.
Similarly, the point with $\tilde{\mu}_H'=\tilde{\mu}_S'$ in the parameter space is critical in the sense that both the $H$ and $S$ directions near their origin become simultaneously flat at the point $\phi=\tilde{\mu}_H'$ ($=\tilde{\mu}_S'$).
Therefore, $\tilde{\mu}_H^2 = \tilde{\mu}_S^2 = 0$ and $\tilde{\mu}_H'=\tilde{\mu}_S'$ is a triple-critical point.
After choosing this point, we may redefine $\phi-\tilde{\mu}_H'$ as a new $\phi$.
This corresponds to the point with $(0,0,0,0,\mu_1,\mu_2,\mu_3)$ in the potential Eq.~(\ref{potential}). 

To obtain a maximal criticality, we need to fix the last three parameters $(\mu_1,\mu_2,\mu_3)$ to a maximal (triple-)critical point. Indeed, the classification of the two-scalar system of $\phi$ and $S$ is given in Ref.~\cite{Kawai:2021lam}.
Among the seven triple-critical points, the critical point {\bf 1234} (CP-1234) is of particular interest because it can realize the FOPT in the early universe without the domain-wall problem.
We thus focus on the CP-1234 in this paper.

\begin{figure}[t]
\begin{center}
\includegraphics[width=10cm]{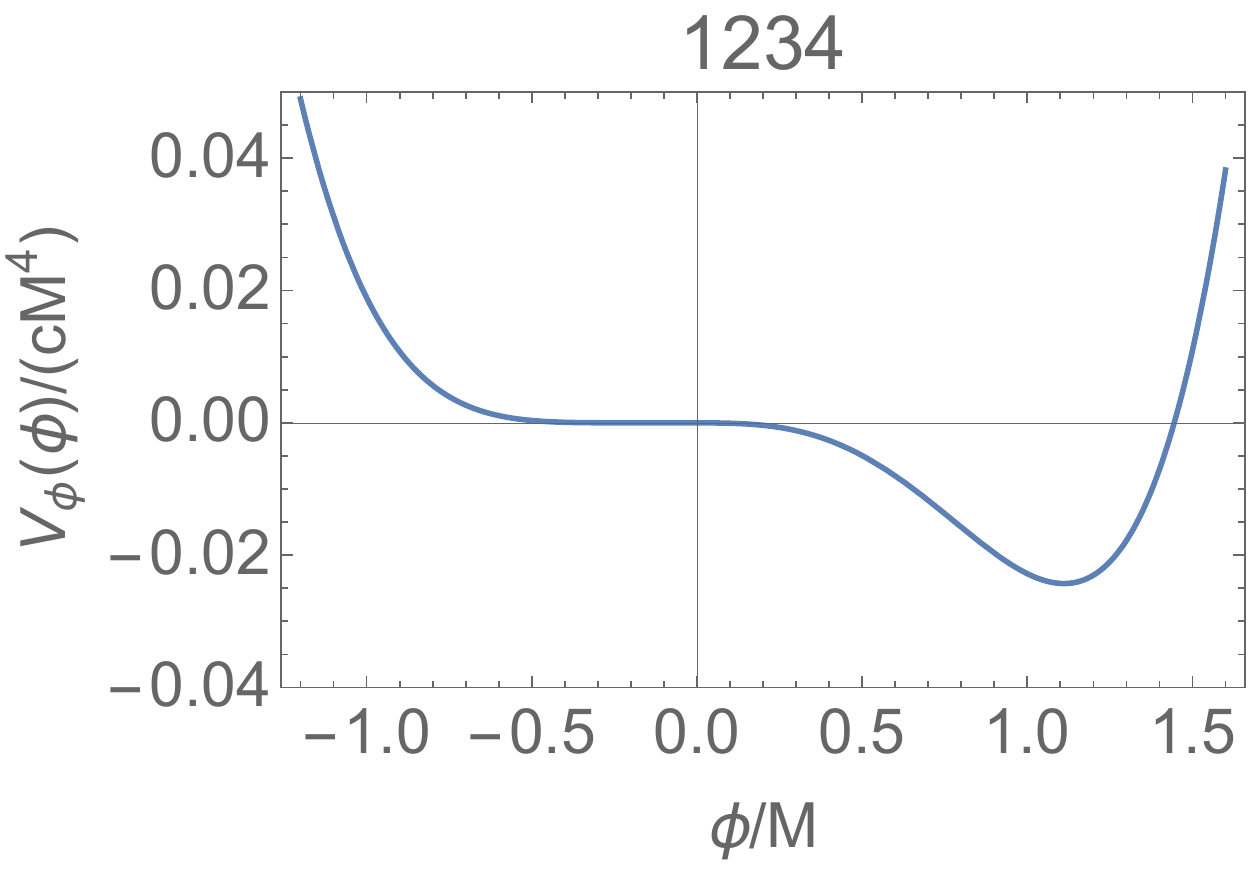}
\caption{
Shape of the potential $V_\phi^{}(\phi)$ normalized by $cM^4$ at the CP-1234. 
}
\label{fig:model}
\end{center}
\end{figure}

We briefly review the CP-1234 and the generalized CW mechanism~\cite{Haruna:2019zeu,Kawai:2021lam}.
The one-loop effective potential with $S= 0$ at zero temperature in the $\overline{\rm MS}$ scheme is given by 
\begin{align}
V_{\rm eff}^{} = & \frac{\lambda_H}{2} (H^\dagger H)^2 - \frac{\lambda_{\phi H}}{2} \phi^2(H^\dagger H)  + \mu_1^{3}\phi + \frac{\mu_2^{2}}{2}\phi^2 + \frac{\mu_3^{}}{3!}\phi^3 +\frac{\lambda_\phi }{4!}\phi^4  \notag\\
&+ \frac{m_\phi^4(\phi)}{64\pi^2}\ln\left(\frac{m_\phi^2(\phi)}{\mu^2 e^{3/2}}\right)  + \frac{m_S^{4}(\phi)}{64\pi^2}\ln \left(\frac{m_S^2(\phi)}{\mu^2e^{3/2}}\right) + \Delta V_\text{1-loop}(h,\mu),
	\label{one-loop potential}
\end{align}
where $\Delta V_\text{1-loop}(h,\mu)$ is the one-loop potential for the physical Higgs $h$ at the renormalization scale~$\mu$, and the effective masses are given by
\aln{
m_\phi^2(\phi)
	&=  \mu_2^{2}+\mu_3^{}\phi + \frac{\lambda_\phi^{}}{2} \phi^2,& m_S^2(\phi)
	&=	\frac{\lambda_{\phi S}^{}}{2}\phi^2.
		\label{effective masses}
}
In Eq.~\eqref{one-loop potential}, we have neglected the $H$-loop contribution to the $\phi$ potential because we focus on the parameter region $\lambda_{\phi H} \ll \lambda_{\phi S}$. Also, the $H$ dependence in Eq.~\eqref{effective masses} are neglected because we consider the region $\lambda_{\phi H}H^\dagger H\ll\lambda_\phi\phi^2$ and $\lambda_{SH}H^\dagger H \ll \lambda_{\phi S}\phi^2$.
We choose the renormalization scale $\mu=\tilde M$ at which $\lambda_\phi=0$~\cite{Kawai:2021lam}:
\aln{
V_{\rm eff}  =&  \frac{\lambda_H}{2} (H^\dagger H)^2 - \frac{\lambda_{\phi H}}{2} \phi^2(H^\dagger H) 
+ \Delta V_{\text{1-loop}}(h,\tilde M)+ \frac{m_\phi^{4}(\phi)}{64\pi^2}\ln\left(\frac{m_\phi^2(\phi)}{\tilde M^{2}e^{3/2}}\right)+V_\phi^{}(\phi)~,\label{Veff2}
}
where
\aln{
V_\phi^{}(\phi)
	&=	\mu_1^{3}\phi+\frac{\mu_2^{2}}{2}\phi^2+\frac{\mu_3^{}}{3!}\phi^3+ \frac{c}{48}\phi^4\ln\left(\frac{\phi^2}{M^2}\right),
		\label{eq:phi_h}
}
in which
\aln{
c 	&= \frac{3\lambda_{\phi S}^2}{16\pi^2},&
M^2	&=	{2\tilde M^{2}e^{3/2}\ov\lambda_{\phi S}}.
}
By neglecting the $m_\phi^{4}(\phi)$ term in Eq.~(\ref{Veff2}), the CP-1234 is realized as 
\aln{
\frac{dV_\phi^{}}{d\phi}\bigg|_{\phi=\phi_S^{}}^{}=\frac{d^2V_\phi^{}}{d\phi^2}\bigg|_{\phi=\phi_S^{}}=\frac{d^3V_\phi^{}}{d\phi^3}\bigg|_{\phi=\phi_S^{}}=\frac{d^4V_\phi^{}}{d\phi^4}\bigg|_{\phi=\phi_S^{}}=0~ 
}
at the field point $\phi=\phi_S^{}$. 
These conditions completely fix the parameters of $V_\phi^{}(\phi)$ as 
\aln{
\mu_1^{3}= -\frac{c}{18}M^3e^{-\frac{25}{4}},~~\mu_2^{2}&= -\frac{c}{4}M^2e^{-\frac{25}{6}},~~\mu_3^{}= -cMe^{-\frac{25}{12}},~~\phi_S= -Me^{-\frac{25}{12}}.
		\label{scalar couplings}
}
Note that $\mu_1^3$, $\mu_2^2$, and $\mu_3$ are proportional to the one-loop suppression factor $c$. This is consistent with having neglected the $m_\phi^{4}(\phi)$ term because $m_\phi^2(\phi)$ is proportional to $c$ as we have chosen the renormalization point such that $\lambda_\phi^{}=0$. 
Substituting Eq.~(\ref{scalar couplings}) into Eq.~(\ref{eq:phi_h}), we obtain 
\begin{align}
V_\phi(\phi)
	&= c M^4 \left[-\frac{\bar{\phi}}{18 e^{25/4}}  -\frac{\bar{\phi}^2}{8e^{25/6}} -\frac{\bar{\phi}^3}{6 e^{25/12}}  + \frac{\bar{\phi}^4}{48}\ln \bar{\phi}^2\right],  &
\bar{\phi}
	&:= \frac{\phi}{M} .\label{eq:vphi2}
\end{align}
The corresponding potential is shown in Fig.~\ref{fig:model}. 
The VEV $v_\phi^{}$ at the true vacuum and the mass of $\phi$ are numerically found to be
\aln{
v_\phi^{}
	&=1.1M,&
m_\phi^2
	&=	-\frac{\lambda_{\phi H}^{}}{2}v^2+0.28cM^2. \label{eq:028}
}

The EW symmetry breaking is triggered by the mixing coupling $\lambda_{\phi H}$, and the Higgs VEV is
\aln{
v := \sqrt{2}\langle H \rangle = v_\phi^{}\sqrt{\frac{\lambda_{\phi H} }{\lambda_H }}~,
\label{ew scale}
}
where $v\simeq 246$~GeV.
In the EW vacuum, the physical scalar states $\hat{h}$ and $\hat{h}_\phi$, which are defined as $\hat{h} = \sqrt{2}\text{Re}H^0 - v$ and $\hat{h}_\phi = \phi - v_\phi$, are mixed with each other. 
The squared mass matrix in the basis of $(\hat{h},\hat{h}_\phi)$ is given by 
\begin{align}
{\cal M}^2 = v^2
\begin{pmatrix} 
\lambda_{H} & -\frac{v}{v_\phi}\lambda_H \\
-\frac{v}{v_\phi}\lambda_H & \frac{m_\phi^2}{v^2}
\end{pmatrix}. 
\end{align}
By introducing the mixing angle $\theta$, the mass eigenstates of the Higgs bosons $(h,h_\phi^{})$ are defined as 
\begin{align}
\begin{pmatrix}
\hat{h} \\
\hat{h}_\phi
\end{pmatrix} = 
\begin{pmatrix}
\cos\theta & -\sin \theta \\
\sin\theta & \cos\theta
\end{pmatrix}
\begin{pmatrix}
h \\
h_\phi
\end{pmatrix}, 
\end{align}
and their squared masses and $\theta$ are expressed as 
\begin{align}
m_h^2      &= {\cal M}^2_{11}\cos^2\theta + {\cal M}^2_{22}\sin^2\theta - {\cal M}^2_{12}\sin 2\theta~, \\
m_{h_\phi}^2 &= {\cal M}^2_{11}\sin^2\theta + {\cal M}^2_{22}\cos^2\theta + {\cal M}^2_{12}\sin 2\theta~, \\
\tan2\theta &= \frac{2{\cal M}^2_{12}}{{\cal M}^2_{11} - {\cal M}^2_{22}}~. 
\end{align}
On the other hand, $S$ does not mix with $\hat{h}$ and $\hat{h}_\phi$ due to the $\mathbb Z_2$ parity, and so its squared mass is simply determined to be 
\begin{align}
m_S^2 = \frac{1}{2}(v^2\lambda_{SH} + v_\phi^2\lambda_{\phi S})~.  \label{eq:mssq}
\end{align}

From the above discussion, two of the six dimensionless parameters $\lambda_H^{}$, $\lambda_\phi^{}$, $\lambda_S^{}$, $\lambda_{\phi S}$, $\lambda_{\phi H}$, and $\lambda_{SH}$ in the potential 
are fixed by $v$ and the Higgs mass $m_h \simeq 125$\,GeV.
For simplicity, we choose $\lambda_S^{}$ to be zero at the EW scale, since it does not play a phenomenologically important role. 
Thus, we can choose the free parameters as 
\aln{
\{m_S ,~\lambda_{SH},~\lambda_{\phi S}^{}\}~~\text{or}~~
\{m_S,~\lambda_{SH},~v_\phi\}.
}
In the latter, the input parameter $\lambda_{\phi S}$ is exchanged by $v_\phi^{}$ via Eq.~(\ref{eq:mssq}). 
Among these couplings, $\lambda_{\phi S}^{}$ is the most important one for the discussion of the FOPT because it determines the finite temperature potential of $\phi$. 

\begin{figure}[t]
\begin{center}
\includegraphics[width=12cm]{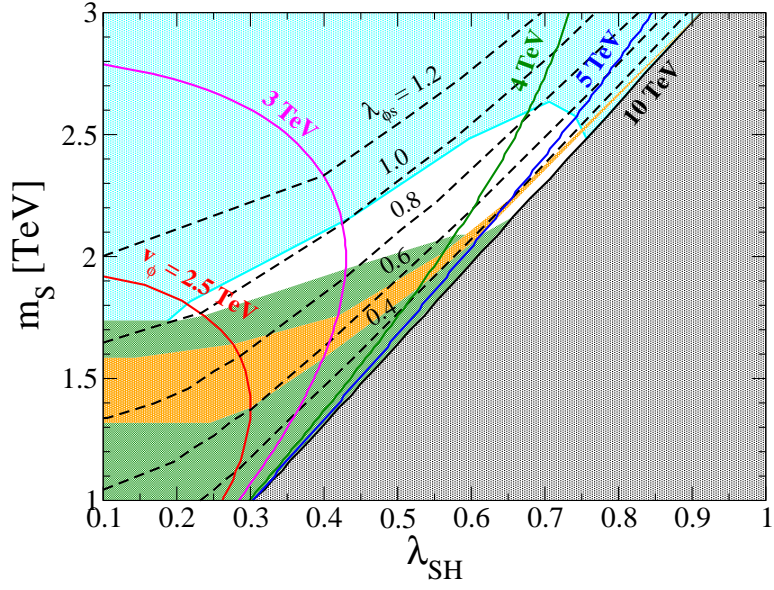}
\caption{
Allowed parameter region (non-shaded) satisfying the observed relic abundance of dark matter on the $\lambda_{SH}$--$m_S^{}$ plane. 
Each of the solid and dashed curves shows the fixed value, written along the curve, of $v_\phi$ and $\lambda_{\phi S}$, respectively. 
Shaded regions are respectively excluded by the XENON1T experiment (green), the LHC data (orange), dark matter abundance (gray), and the perturbativity bound (cyan).
Regarding the perturbativity bound, the absence of Landau pole up to $\mu=10^{17}$ GeV is imposed.  
}
\label{fig:const}
\end{center}
\end{figure}

As aforementioned, $S$ is the dark matter candidate in this model. 
Its thermal relic abundance and the constraint from direct searches at XENON1T~\cite{Aprile:2018dbl} have been investigated in Ref.~\cite{Hamada:2020wjh} for some of double-critical points in the $(\mu_1^3,\mu_2^2,\mu_3)$ space.
On the other hand, we study the maximal (triple-)critical point in the current paper.
The differences in the predictions of $m_{h_\phi}^{}$ and $\theta$ from the previous analysis solely originate from the numerical factor 0.28 in Eq.~(\ref{eq:028}).
The current maximally critical model tends to predict a slightly larger mass of $h_\phi^{}$ compared with the previous one.  
This makes the bound from the direct search slightly weaker. 
On the other hand, the relic abundance is well fitted by the following formula~\cite{Hamada:2020wjh}:
\aln{
4\lambda_{SH}^2+\lambda_{\phi S}^2
	&=	\left(\frac{m_S }{m_{\rm th} }\right)^2, &
m_{\rm th}
	&=	1590\pm40~\text{GeV}. 
\label{eq:fitting}
} 
We note that in the numerical evaluation of the relic abundance and the constraint from the direct search, 
we employ the {\tt MicrOMEGAs} package~\cite{Belanger:2018ccd}, 
and impose the relic abundance to be $\Omega_S h^2 = 0.12$~\cite{Aghanim:2018eyx}.

In Fig.~\ref{fig:const}, we show the regions allowed (non-shaded area) by various constraints on the $\lambda_{SH}$--$m_S^{}$ plane. 
Each point in the allowed region satisfies the relic abundance, i.e., $v_\phi^{}$ and $\lambda_{\phi S}$ are determined such that they satisfy Eqs.~(\ref{eq:mssq}) and (\ref{eq:fitting}).
The solid and dashed curves respectively show the contours of fixed values of $v_\phi^{}$ and $\lambda_{\phi S}$, where 
the number is written along each curve.  
We also show the other constraints on the model parameters in the figure: 
The green, orange, gray, and cyan shaded regions are excluded by the updated XENON1T result~\cite{Aprile:2018dbl},
\footnote{
For the region with the dark matter mass larger than 1 TeV, we extrapolate the upper limit on the spin-independent cross-section for dark matter and nucleon scatterings.} the LHC data, dark matter relic abundance, and perturbativity bound, respectively~\cite{Hamada:2020wjh}.
For the LHC bound, we demand that the signal strength of the Higgs boson $\mu_h$ is given within the 2$\sigma$ range, where the current measurement with the integrated luminosity of 139 fb$^{-1}$ 
shows $\mu_h = 1.06\pm 0.06$~\cite{ATLAS:2021vrm}.  
In our scenario, $\mu_h$ can differ from unity due to the mixing with the singlet-like Higgs boson $h_\phi$, so this constraint can be described as $\sin^2\theta \leq 0.06$. 
We see that the successful benchmark scenario is given at the well constrained region, i.e., 
$1.7\,\text{TeV}\lesssim m_S \lesssim 2.5$\,TeV, 
$0.2 \lesssim \lambda_{SH} \lesssim 0.75$, and  
$0.4 \lesssim \lambda_{\phi S} \lesssim 0.8$. 
In Table~\ref{tab:benchmark}, we show the range of parameters allowed by all the constraints discussed above for each fixed value of $v_\phi$. 

\begin{table}[!t]
\begin{center}
\begin{tabular}{cccccccccccccc}\hline\hline
  $v_\phi$ [TeV] & $\lambda_{SH}$  & $\lambda_{\phi S}$  & $m_S^{}$ [GeV] & $m_{h_\phi}^{}$ [GeV] & $\sin\theta$ \\\hline
  2.5            & (0.21,0.24)  & (0.99,1.1)   & (1760,1810)    & (160,170) & (0.11,0.13) \\\hline
  3              & (0.42,0.43)  & (0.84,1.0)   & (1950,2140)    & (170,200) & (0.052,0.10) \\\hline
  4              & (0.58,0.68)  & (0.54,0.84)  & (2090,2600)    & (140,220) & (0.029,0.18) \\\hline
  5              & (0.66,0.74)  & (0.40,0.53)  & (2250,2570)    & (140,170) & (0.053,0.24) \\\hline
  10             & (0.65,0.76)  & (0.093,0.12) & (2160,2480)    & (60,80) & ($-$0.042,$-$0.032) \\\hline\hline
\end{tabular}
\caption{Range of parameters allowed by all the constraints.  }
\label{tab:benchmark}
\end{center}
\end{table}

\section{First Order Phase Transition}\label{sec:FOPT}

In this section, we estimate the bubble nucleation and percolation temperatures of the FOPT using the effective potential of $\phi$ at finite temperature.  
For the model described in Section~\ref{Model}, we will see that the potential barrier disappears below $T=0.26\,T_\text{c}^{}$ due to the existence of the linear term in the zero-temperature effective potential~(\ref{Veff2}), where $T_\text{c}$ is the critical temperature at which the free energies of two vacua become degenerate. This implies that the percolation temperature cannot fall below $0.26\,T_\text{c}^{}$.  
In Section~\ref{FTP}, we study the effective potential of $\phi$ at finite temperature. 
We then discuss the nucleation and percolation in Section~\ref{nucleation and percolation}. 
\subsection{Finite temperature potential}\label{FTP}
\begin{figure}[t]
\begin{center}
\includegraphics[width=8.5cm]{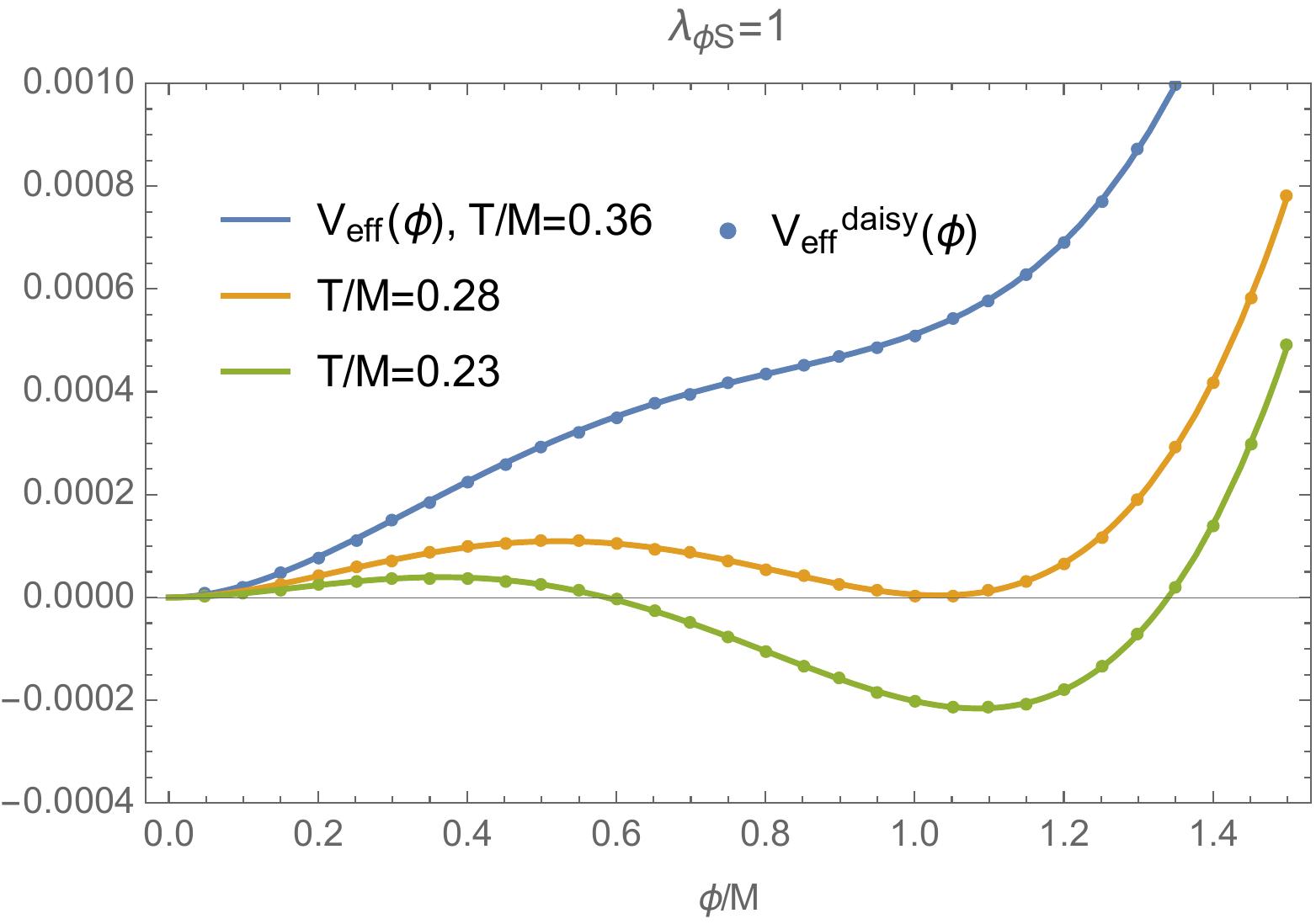}
\includegraphics[width=8.5cm]{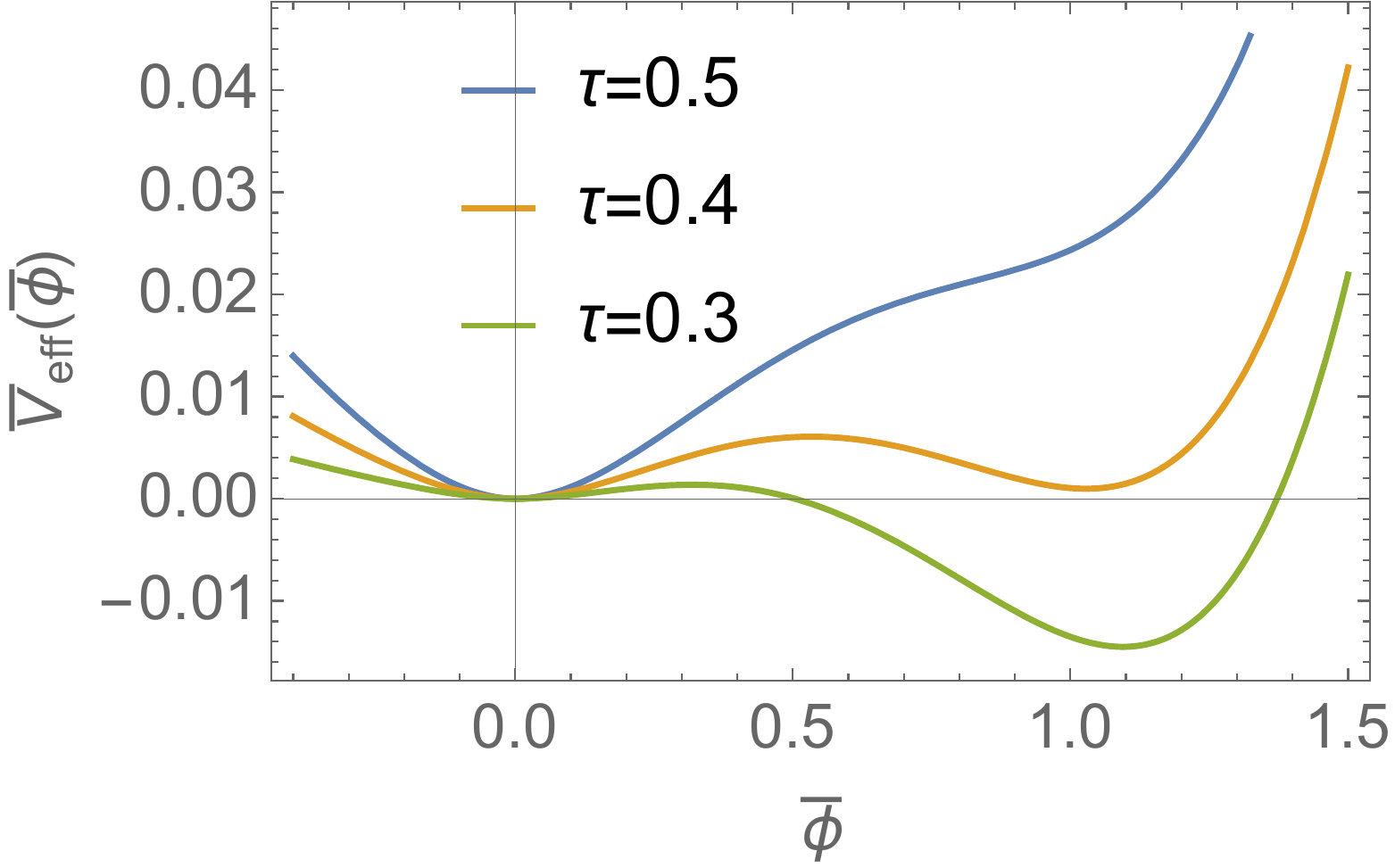}\\ \vspace{5mm}
\includegraphics[width=10cm]{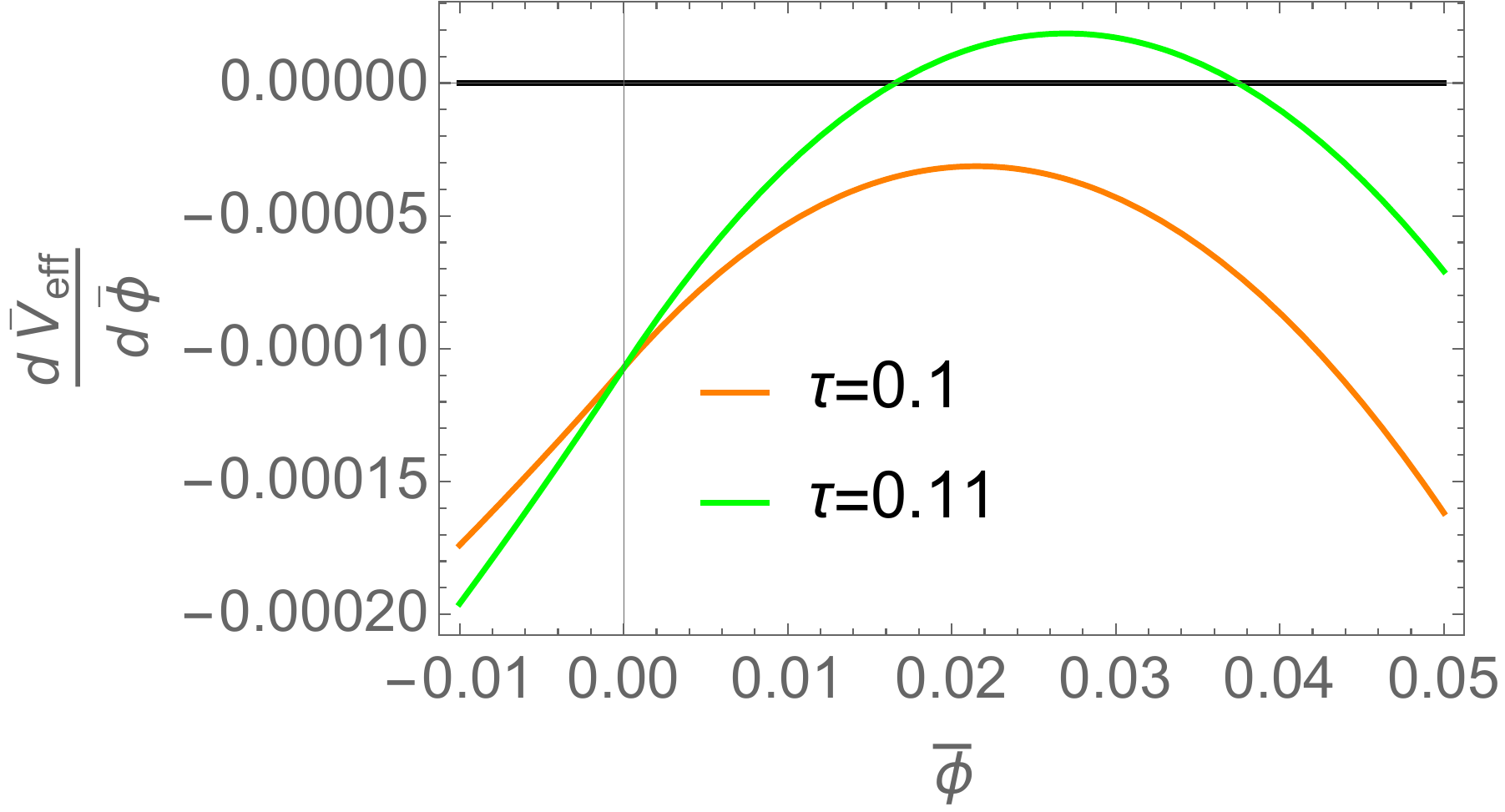}
\end{center}
\caption{Upper left: Comparison between $V_{\rm eff}^{}(\phi,T)$ (curves) and $V_{\rm eff}^{\rm daisy}(\phi,T)$ (dots).
Upper right: Normalized effective potential $\overline{V}_{\rm eff}(\overline{\phi};\tau)$. 
Lower: The $\overline{\phi}$-derivative of $\overline{V}_{\rm eff}(\overline{\phi};\tau)$. 
}
\label{fig:daisy}
\end{figure}

Let us study the one-loop effective potential at finite temperature. 
In general, the one-loop thermal correction is 
\aln{
 V^\text{1-loop}_{\rm th}(\phi,T)
= \frac{T^4}{2\pi^2}\sum_i \left[g_{\text Bi}^{}I_\text{B}^{}\left(\frac{m_{\text Bi}(\phi)}{T}\right) - g_{\text Fi}^{}I_\text{F}^{}\left(\frac{m_{\text Fi}(\phi)}{T}\right)\right]~,
}
where $g_{\text Bi}^{}$ and $m_{\text Bi}$~($g_{\text Fi}^{}$ and $m_{\text Fi}$) represent the degrees of freedom and mass for boson (fermion), respectively, and the functions $I_\text{B}$ and $I_\text{F}$ are given by 
\aln{
I_{B\atop F}(a) & = \int_0^\infty dxx^2\log\left(1 \mp e^{-\sqrt{x^2 + a^2}}\right). 
}
In principle, all the scalar fields may contribute to the one-loop potential of $\phi$.  
However, the contributions from $H$ and $\phi$ are smaller than from $S$ because $m_H^2(\phi)\ll m_S^2(\phi)$ and $m_\phi^2(\phi)\ll m_S^2(\phi)$ as discussed below Eqs.~\eqref{effective masses} and \eqref{scalar couplings}.
Thus, it suffices to take into account the $S$ contribution only:\footnote{
In fact, the background mass of $\phi$ is $m_\phi^{2}(\phi)=\mu_2^{2}+\mu_3^{}\phi/3\propto \lambda_{\phi S}^2/16\pi^2$, from which we can estimate the mass ratio at around $\phi=M$ as $m_\phi^2(M)/m_S^2(M)\sim \lambda_{\phi S}/16\pi^2\lesssim 0.06$ for $\lambda_{\phi S}\lesssim 1$. Thus, the contribution from $\phi$ is at most $10$\% of that from $S$. 
}
%
\aln{
V_{\rm eff}^{}(\phi,T)
	&=	V_\phi^{}(\phi)+\frac{T^4}{2\pi^2}I_\text{B}\left(\frac{m_{S}(\phi)}{T}\right) ,&
\quad m_S^{2}(\phi)
	&=	\frac{\lambda_{\phi S}^{}}{2}\phi^2.
}
At high temperatures, the perturbative calculations may break down, and it is necessary to sum up all the relevant diagrams~\cite{Curtin:2016urg,Senaha:2020mop}. 
The so-called daisy resummation in the Parwani scheme~\cite{Parwani:1991gq} can be obtained by simply replacing $m_S(\phi)$ with the thermal corrected effective mass $m_S^{\rm daisy}(\phi,T)$:
\aln{
V_{\rm eff}^{\rm daisy}(\phi,T)=V_\phi^{}(\phi)+\frac{T^4}{2\pi^2}I_\text{B}\left(\frac{m_{S}^{\rm daisy}(\phi,T)}{T}\right)~,
}
where
\aln{
\left(m_S^{\rm daisy}(\phi,T)\right)^2&=m_S^{2}(\phi)+\frac{\lambda_{\phi S}^{}T^2}{2}J\left(\frac{m_S^{2}(\phi)}{T^2}+\frac{\lambda_{\phi S}^{}}{24}\right)~,
\\
J(\alpha)&:=\frac{1}{\pi^2}\frac{\partial}{\partial \alpha}I_\text{B}^{}(\alpha)=\frac{1}{12}-\frac{\alpha^{1/2}}{4\pi}+\cdots~.
}
In the upper-left panel in Fig.~\ref{fig:daisy}, we show $V_{\rm eff}^{}(\phi,T)$ (curves) and $V_{\rm eff}^{\rm daisy}(\phi,T)$ (dots) for different values of $T$. 
Although there actually exist tiny differences between these potentials, the difference is less than 1$\%$ even when $T/T_\text{c}^{}\sim 1$ ($T_\text{c}^{}$ is $\sim 0.28M$ from the figure).  
Therefore, we use $V_{\rm eff}^{}(\phi,T)$ in the following discussion. 

To understand the parameter dependence of the FOPT, it is convenient to 
normalize the effective potential by $cM^4$:
\aln{\overline{V}_{\rm eff}^{}(\overline{\phi},\tau)
	&:= 
\frac{V_{\rm eff}^{}(\phi,T)}{cM^4}
	=
-\frac{\overline{\phi}}{18e^{25/4}}-\frac{\overline{\phi}^2}{8e^{25/6}}-\frac{\overline{\phi}^3}{6e^{25/12}}+\frac{\overline{\phi}^4}{48}\log \overline{\phi}^2
+\frac{2\tau^4}{3}I_\text{B}\left(\frac{\bar{\phi}}{\tau} \right),&
\tau &:= {T\over m_S^{}(M)}.
}
One can see that the parameter dependence of the effective potential appears only through $\tau$ except for the overall normalization so that the critical behavior is determined only by the $\tau$ parameter. 
In the upper right panel in Fig.~\ref{fig:daisy}, we plot $\overline{V}_{\rm eff}^{}$ where the different colors correspond to the different values of $\tau$. 
The critical value of $\tau$, i.e., $\tau_\text{c}$ at which the two vacua are degenerate
is numerically found to be 
\aln{
\tau_\text{c}^{}=0.39~, \label{bc}
} 
by which the critical temperature $T_\text{c}$ is determined to be
\aln{
T_\text{c}^{}=0.39\, m_S^{}(M),
}
and hence $\tau=0.39\,T/T_\text{c}^{}$.  

There is another important value of $\tau$, denoted by $\tau_\text{b}$, at which the potential barrier disappears due to the linear term of the potential $V_\phi^{}$ at zero temperature.
The value of $\tau_\text{b}$ is numerically found to be
\aln{\tau_\text{b}^{} = 0.10~, \label{bm}
}
which corresponds to
\aln{
T_\text{b}^{}=0.10 \, m_S^{}(M)\simeq 0.26\,T_\text{c}^{}~.
}

For illustration, in the lower panel of Fig.~\ref{fig:daisy}, $d\overline{V}_{\rm eff}/d\overline{\phi}$ is shown for $\tau=0.1$ (orange) and $\tau=0.11$ (green).
When $\tau=0.11$, the derivative of the potential is always negative, which means that there is no potential barrier in this case.
When temperature is $T_\text{n}\lesssim T\lesssim T_\text{c}$, the universe is supercooled, where $T_\text{n}$ is the bubble nucleation temperature. 
In the present case however, there is not large room for the supercooling because $T_\text{b}=0.26\,T_\text{c}$. 
%


\subsection{Nucleation and Percolation}\label{nucleation and percolation}

We estimate the bubble nucleation and percolation temperatures. 
The dynamics of phase transition is determined by the decay rate of the false vacuum into the true vacuum per unit time and volume:
\aln{
\Gamma(T)\sim T^4\exp\left(-\frac{S_3^{}(T)}{T}\right)~, \label{eq:gamma}
}
where $S_3$ is the $O(3)$-symmetric bounce action. 
The bubble nucleation is realized when 
\begin{align}
\frac{\Gamma(T_\text{n})}{\bigl(H(T_\text{n})\bigr)^4} \sim 1,
\label{Gamma over Hubble 4}
\end{align}
where $H(T)$ is the Hubble scale at temperature $T$. 
In the radiation-dominated era, $H(T)$ is given by (see e.g.\ Ref.~\cite{Kolb:1990vq})
\begin{align}
H(T) = 1.66\sqrt{g_*}\frac{T^2}{M_\text{P}}, 
\end{align}
where $g_* \sim 100$ is the effective number of massless degrees of freedom and $M_\text{P}=1.2\times10^{19}\,\text{GeV}$ is the Planck mass. 
To estimate the nucleation temperature $T_\text{n}$, we evaluate $S_3$ at temperature $T$ given by 
\aln{
S_3^{}=\frac{4\pi M}{\sqrt{c}}\int_0^\infty d\rho\rho^2\left[\frac{1}{2}\left(\frac{d\overline{\phi}_\text{B}}{d\rho}\right)^2+\overline{V}_{\rm eff}^{}(\overline{\phi}_\text{B};\tau)\right], 
\label{O(3) action}
}
where $\rho$ is the normalized radial coordinate of the 3-dimensional Euclidean space $r$ defined by $\rho = c^{1/2}M r$. 
In Eq.~(\ref{O(3) action}), $\bar{\phi}_\text{B}$ is the bounce solution of $\bar{\phi}$ satisfying the following equation of motion:
\aln{
\frac{d^2\overline{\phi}_\text{B}}{d\rho^2}+\frac{2}{\rho}\frac{d\overline{\phi}_\text{B}}{d\rho}
	&=	\frac{\partial \overline{V}_{\rm eff}^{}}{\partial \overline{\phi}}\Big|_{\overline{\phi}=\overline{\phi}_\text{B}},&
\overline{\phi}_\text{B}(\infty)
	&=	\overline\phi_\text{FV},&
\frac{d \overline{\phi}_\text{B}}{d\rho}\bigg|_{\rho=0}
	&=0,
}
where $\overline\phi_\text{FV}$ is the location of the false vacuum near the origin, which is numerically obtained for each $\tau$.

After the bubble nucleation, there is a temperature at which true vacuum bubbles start to form an infinite connected cluster. 
%
Such a temperature is referred to as the percolation temperature $T_\text{p}$, which is customary estimated by~\cite{Wang:2020jrd}
\aln{
I(T_\text{p}^{}) \sim  0.34~,
	\label{Tp determination}
}
where $I(T)$ is the exponent of the probability of finding a point still in the false vacuum: $p(T)=e^{-I(T)}$.
It is given by 
\aln{
I(T)
	&=	\int_{t_\text{c}^{}}^tdt' \Gamma(t') a(t')^3\times \frac{4\pi}{3}R(t,t')^3~, & 
R(t,t')
	&=	v_\text{w}^{}\int_{t'}^t\frac{ds}{a(s)}~, 
}
where $a(t)$ and $v_\text{w}^{}$ are the scale factor at time $t$ and the bubble wall velocity, respectively. 
The determination of $v_\text{w}^{}$ is one of the challenging issues in the field of FOPTs~\cite{Gouttenoire:2021kjv}. 
We choose $v_\text{w}^{}=0.5$~\cite{Caprini:2015zlo,Caprini:2019egz} as a benchmark value in this paper.    
For numerical evaluations, we determine $T_\text{n}^{}$ and $T_\text{p}$ by assuming that Eqs.~(\ref{eq:gamma}), \eqref{Gamma over Hubble 4}, and \eqref{Tp determination} exactly hold.
When $g_* = 100$ and $M_\text{P}/T_\text{n}=10^{17}$, the condition for $T_\text{n}$ can be expressed as $S_3/T_\text{n} \simeq 145$. 

\begin{figure}[t]
\begin{center}
\includegraphics[width=8cm]{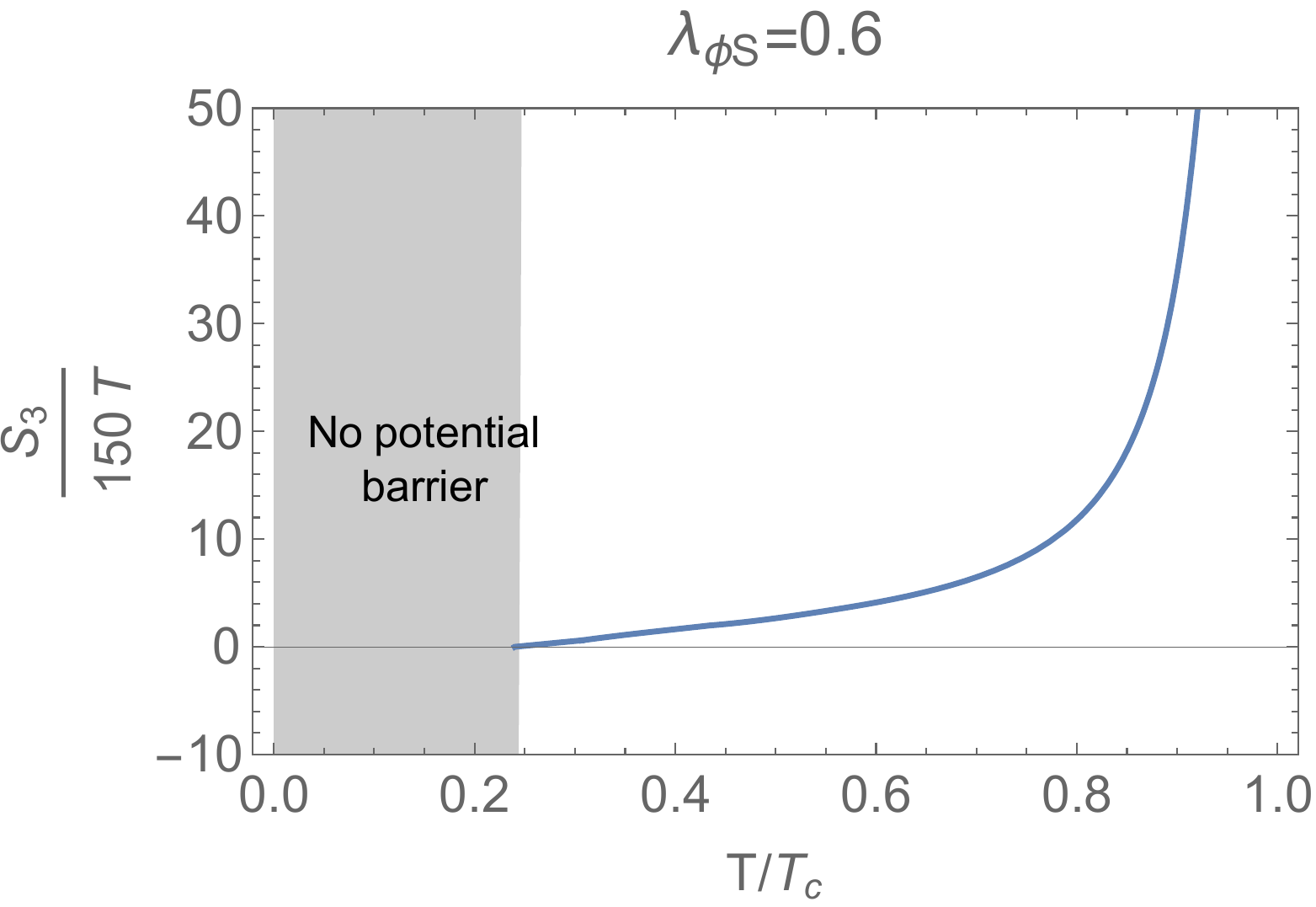}
\includegraphics[width=8cm]{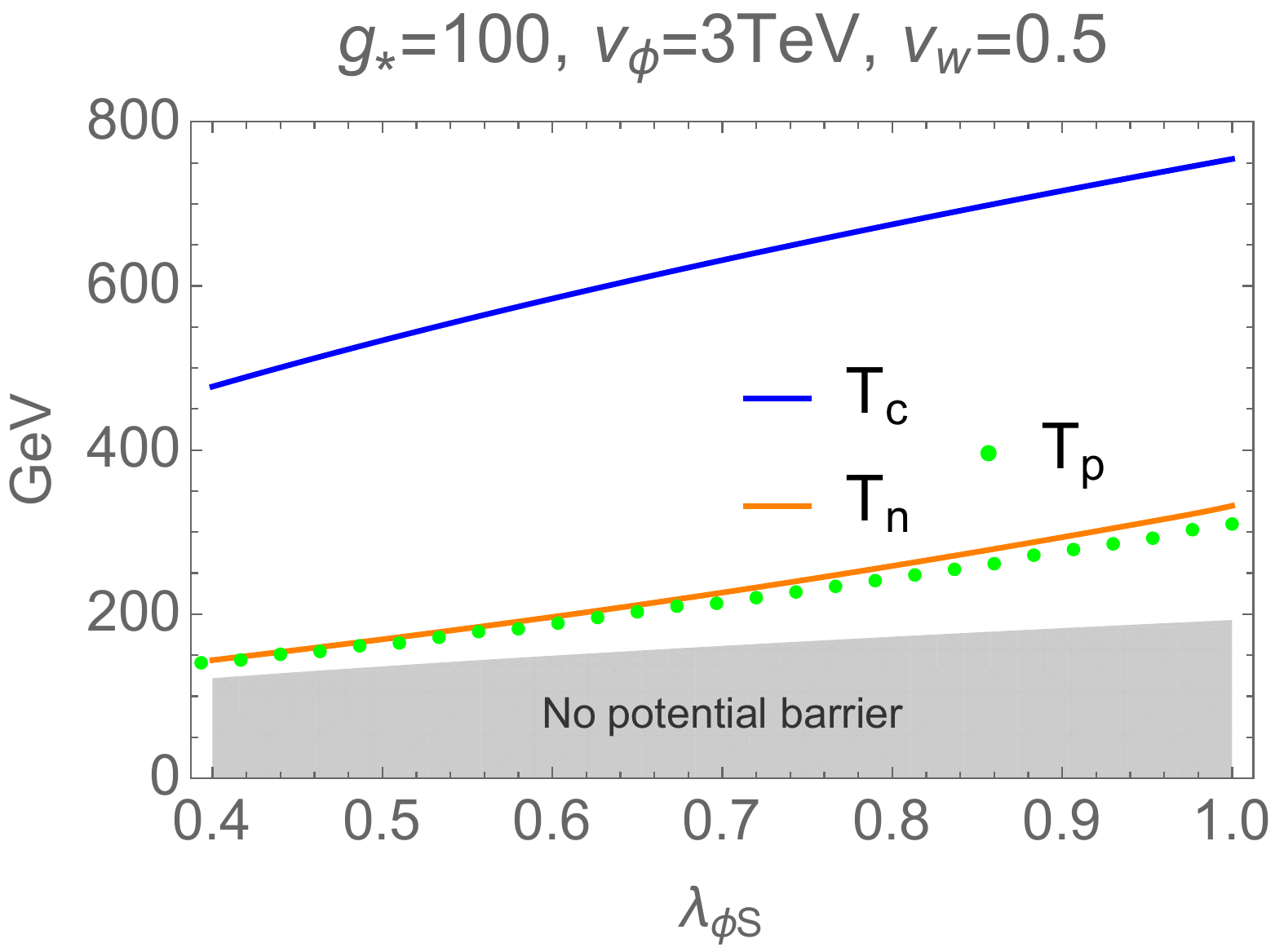}
\end{center}
\caption{Left: Bounce action $S_3^{}(T)/T$. Right: Various temperatures: The blue and orange lines are the critical and nucleation temperatures $T_\text{c}^{}$ and $T_\text{n}^{}$, respectively, and the green points are the percolation temperature $T_\text{p}^{}$. In both panels, gray region is $T<T_\text{b}=0.25T_\text{c}$.
}
\label{fig:bounce}
\end{figure}

In the left panel in Fig.~\ref{fig:bounce}, we plot the numerical values of $S_3^{}(T)/T$ for $\lambda_{\phi S}^{}=0.6$ 
by using the {\tt FindBounce} package~\cite{Guada:2020xnz}\footnote{As a consistency check, we have also calculated the bounce action by using the {\tt  CosmoTransitions} \cite{Wainwright:2011kj} and obtained the same results within a few percent accuracy. 
}. 
Here, the gray region represents $\tau<\tau_\text{b}$ ($= 0.10$) $\leftrightarrow T/T_\text{c}^{}<0.25$, where the potential barrier no longer exists.  

In the right panel in Fig.~\ref{fig:bounce}, we plot $T_\text{c}^{}$ (blue), $T_\text{n}^{}$ (orange), and $T_\text{p}^{}$ (green points) as functions of $\lambda_{\phi S}^{}$ for $v_\phi^{}=3~$TeV and $v_\text{w}^{}=0.5$.  
As in the left panel, the potential barrier disappears in the gray region. 
One can see that there is no significant difference between $T_\text{n}^{}$ and $T_\text{p}^{}$ in the present model, which means that the percolation occurs soon after the onset of nucleation. 
%
Note also that $T_\text{n}$ and $T_\text{p}$ become smaller than $T_\text{b}^{}$ for $\lambda_{\phi S}^{}\lesssim 0.4$. We have checked that this is the case for $3~{\rm TeV}\lesssim  v_\phi^{}\lesssim 10~{\rm TeV}$.
In particular for $v_\phi^{}=10~$TeV, a FOPT cannot be realized consistent with other phenomenological constraints because the allowed region of $\lambda_{\phi S}^{}$ is smaller than $0.4$  (see Table~1).

\section{Gravitational Wave Signals}\label{sec:GW}
In this section, we study the GW signals produced by the FOPT.  
In general, there are three kinds of sources of the stochastic GWs produced during a FOPT: bubble collisions, sound waves, and magnetohydrodynamic (MHD) turbulence in the plasma~\cite{Caprini:2015zlo,Caprini:2019egz}. 
Typically, the dominant source is the sound waves as long as the FOPT is not too strong, i.e., $\alpha\lesssim 1$ and $\beta/H\gtrsim 100$ in the following notation.

\subsection{Strength parameters}
\begin{figure}[t]
\begin{center}
\includegraphics[width=5cm]{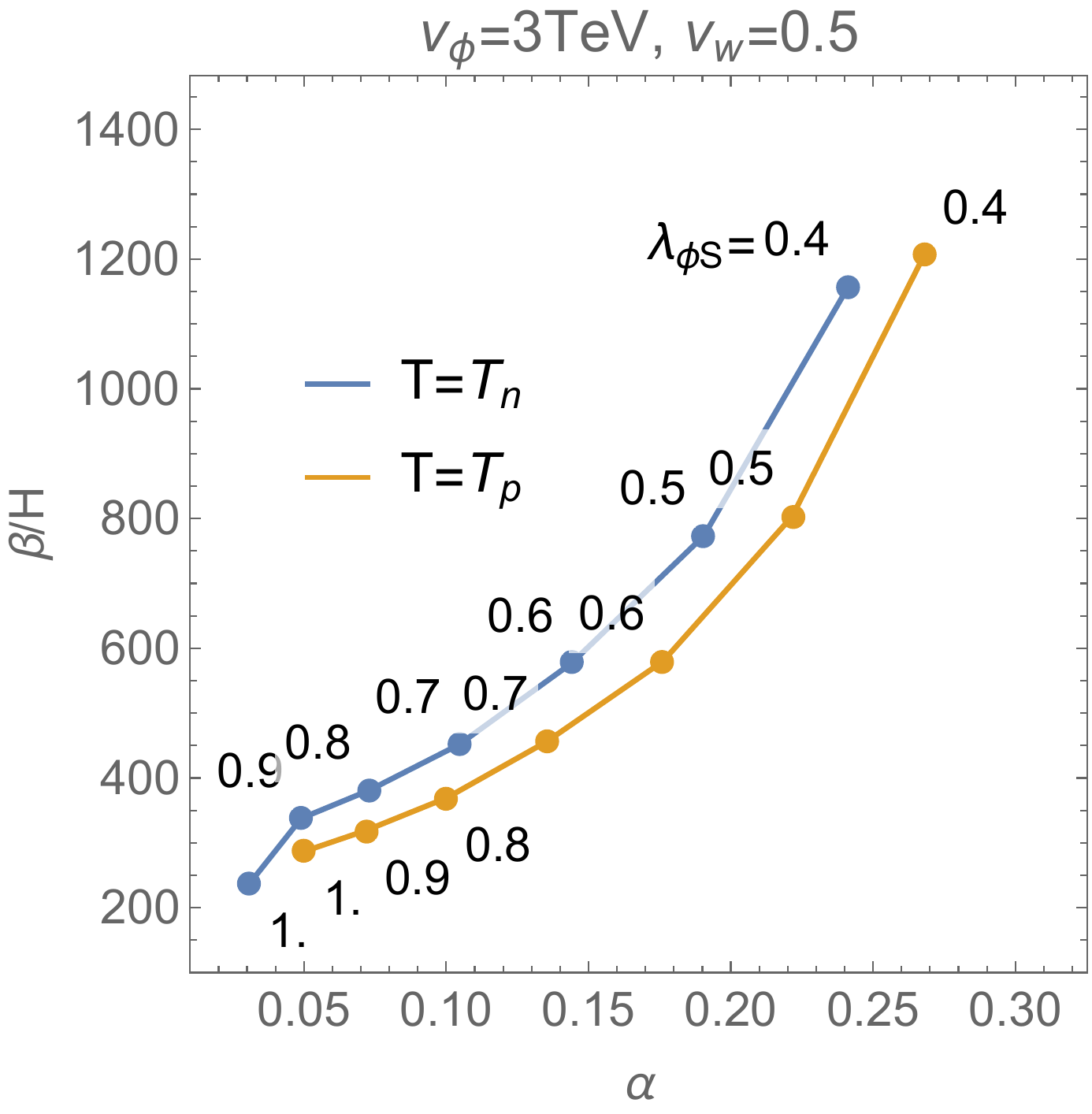}
\includegraphics[width=5cm]{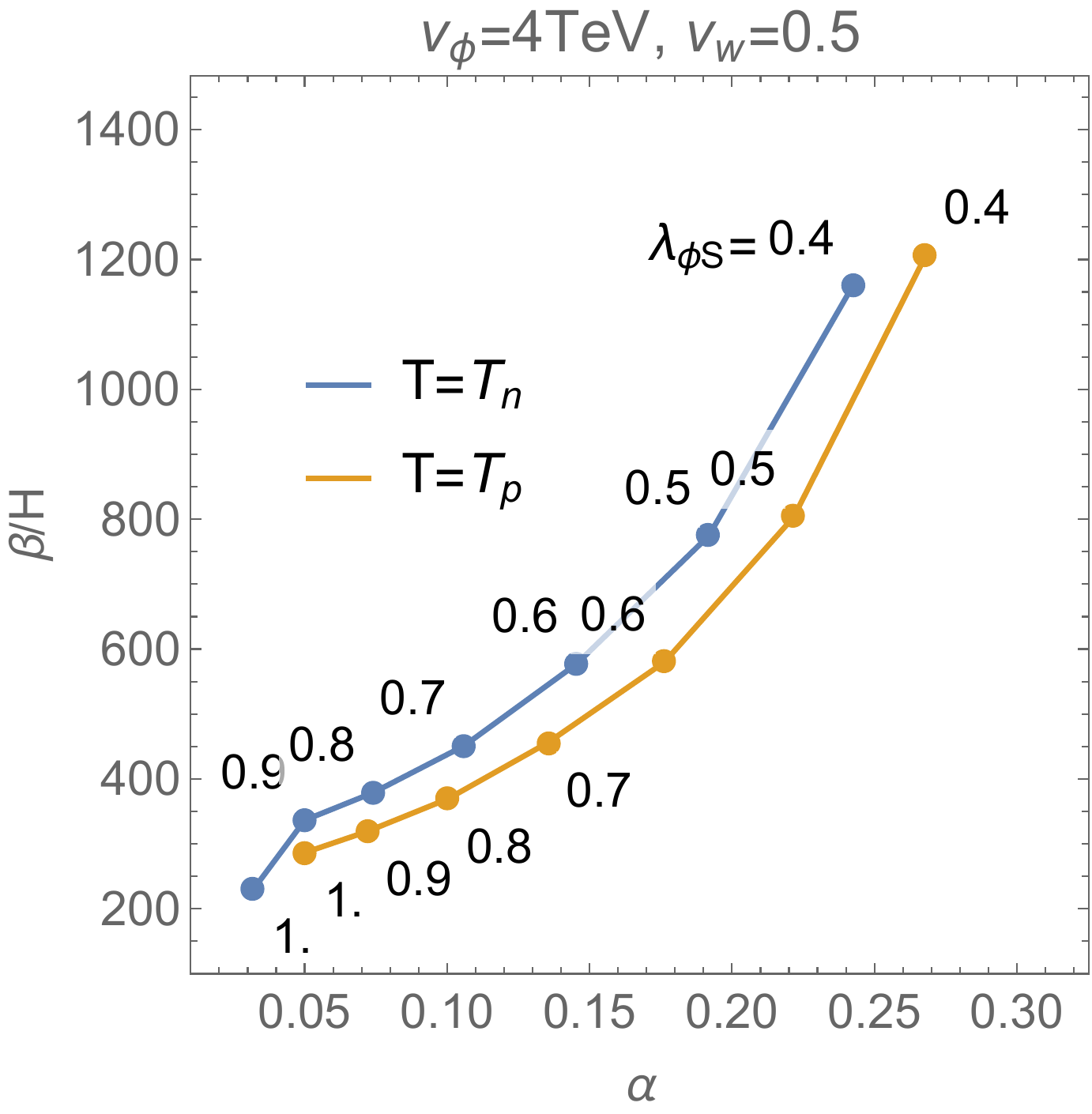}
\includegraphics[width=5cm]{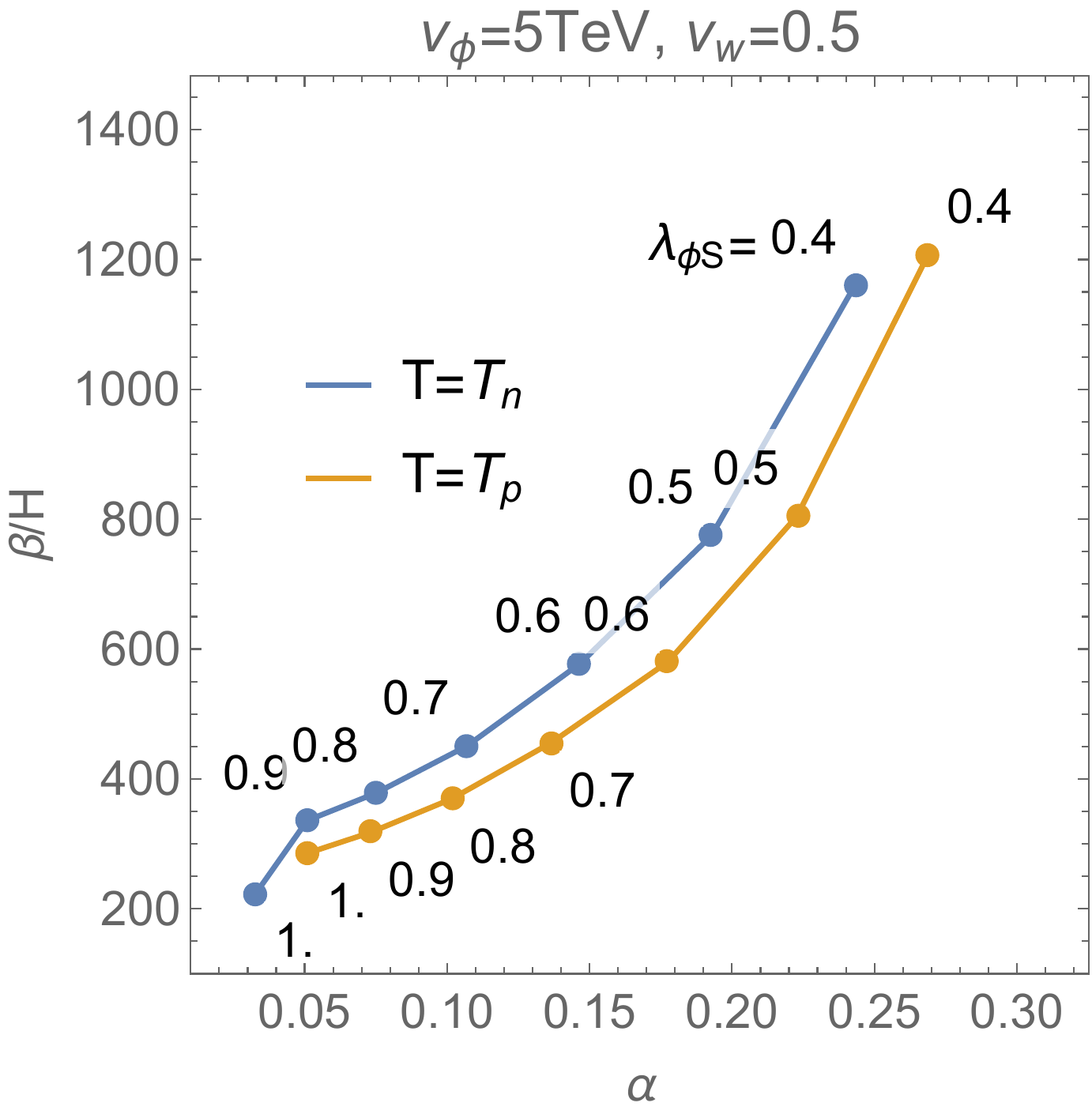}
\end{center}
\caption{Predictions of $\alpha$ and $\beta/H$ for $v_\phi^{}=3~$TeV (left), $v_\phi^{}=4~$TeV (middle), and $v_\phi^{}=5~$TeV (right). 
}
\label{fig:alpha-beta}
\end{figure}
There are two important parameters, $\alpha$ and $\beta$, to determine the GW energy spectrum $\Omega_{\rm GW}^{}(f)$ produced by a FOPT. 
The parameter $\alpha$ measures the strength of the FOPT, which is defined by the ratio between the latent heat energy and the radiation energy $\rho_\text{R}$:
\aln{\alpha(T):= \frac{1}{\rho_\text{R}(T)}\left[\Delta V_{\rm eff}^{}(T)+T\frac{\partial \Delta V_{\rm eff}^{}(T)}{\partial T}\right]~,
\label{alpha}
}
where
\aln{\Delta V_{\rm eff}^{}(T)=V_{\rm eff}^{}(\phi_{\rm false}^{},T)-V_{\rm eff}^{}(\phi_{\rm true}^{},T)~.
}
We note that the second term in Eq.~(\ref{alpha}), i.e., the contribution from the difference of entropy, is negligibly smaller than the first term. 
The other parameter $\beta$ is defined by
\aln{\beta(T):=-H(T)T\frac{\partial \ln \Gamma (T)}{\partial T}~,
}
which determines the duration of the FOPT and the characteristic frequency of the GWs.    
In the following, we will consider two commonly assumed options $T=T_\text{p}^{}$ and $T=T_\text{n}^{}$ for the typical temperature to evaluate $\alpha$ and $\beta$~\cite{Caprini:2015zlo,Jinno:2016knw,Caprini:2019egz}. 

\

In Fig.~\ref{fig:alpha-beta}, we show our numerical results in the $\alpha$-$\beta$ plane for $v_\phi^{}=3~$TeV (left), $v_\phi^{}=4~$TeV (middle), and $v_\phi^{}=5~$TeV (right), where $\lambda_{\phi S}^{}$ is varied in each contours within the range $0.4\leq \lambda_{\phi S}^{}\leq 1$.  
The orange and blue colors correspond to the choices $T=T_\text{p}^{}$ and $T=T_\text{n}^{}$, respectively.
The orders of magnitude of $\alpha$ and $\beta/H$ are $10^{-2}\text{--}10^{-1}$ and $10^{2}\text{--}10^{ 3}$, respectively, indicating that the FOPT is moderate in this model. 
In particular, the large $\beta/H$ corresponds to the short duration of the FOPT, and this is why $T_\text{n}^{}$ and $T_\text{p}^{}$ are close to each other as seen in the previous section.    

\

To obtain the GW spectrum, we also need to calculate the energy efficiency factor $\kappa_\text{v}^{}$ which is defined as the fraction of vacuum energy that is transferred to the bulk motion of plasma. 
In this paper, we use the analytical fitting result presented in Ref.~\cite{Espinosa:2010hh}:
\aln{\kappa_\text{v}^{}
=\frac{c_\text{s}^{11/5}\kappa_A^{}\kappa_\text{B}^{}}{(c_\text{s}^{11/5}-v_\text{w}^{11/5})\kappa_\text{B}^{}+v_\text{w}^{}c_\text{s}^{6/5}\kappa_A^{}}
\quad
(\text{for $v_\text{w}^{}\lesssim c_\text{s}^{}=1/\sqrt{3}$})~,
}   
where $c_\text{s}\sim 0.58$ is the speed of sound and  
\aln{\kappa_A^{}
	&=	v_\text{w}^{6/5}\frac{6.9\alpha}{1.36-0.037\sqrt{\alpha}+\alpha}~,&
\kappa_\text{B}^{}
	&=	\frac{\alpha^{2/5}}{0.017+(0.997+\alpha)^{2/5}}~.
}

\subsection{Gravitational Wave Signals}
\begin{figure}[t]
\begin{center}
\includegraphics[width=8cm]{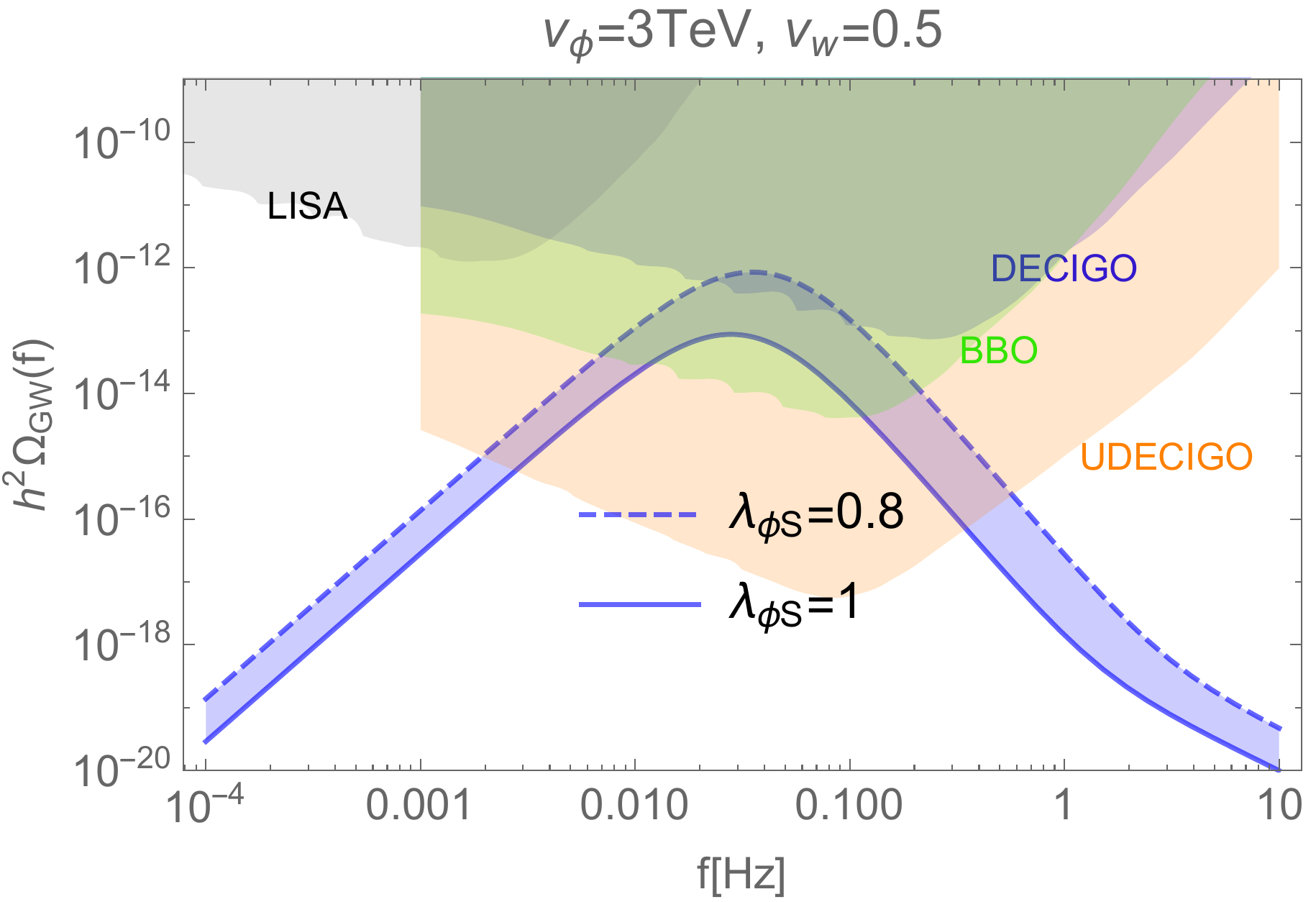}
\includegraphics[width=8cm]{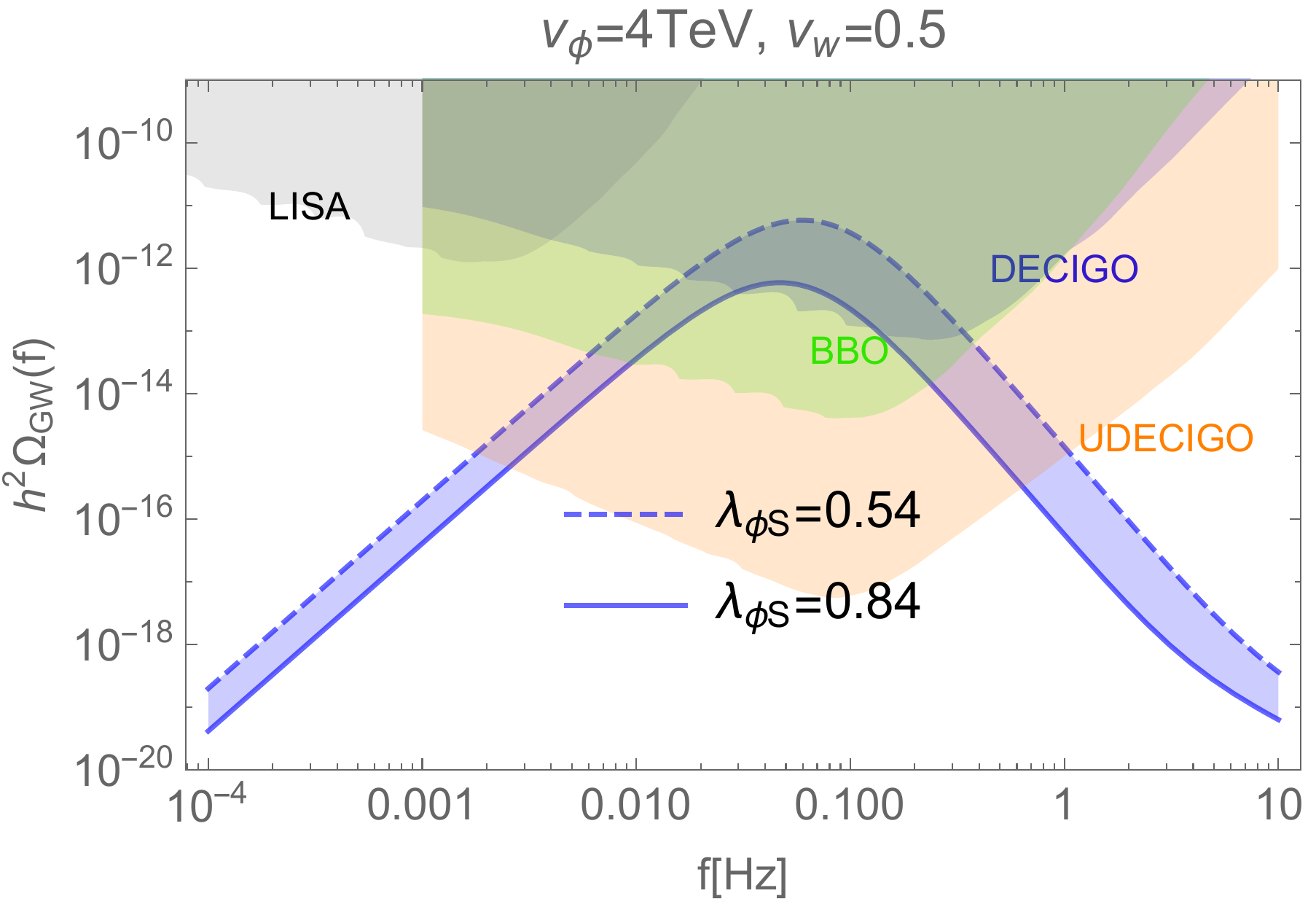}
\includegraphics[width=8cm]{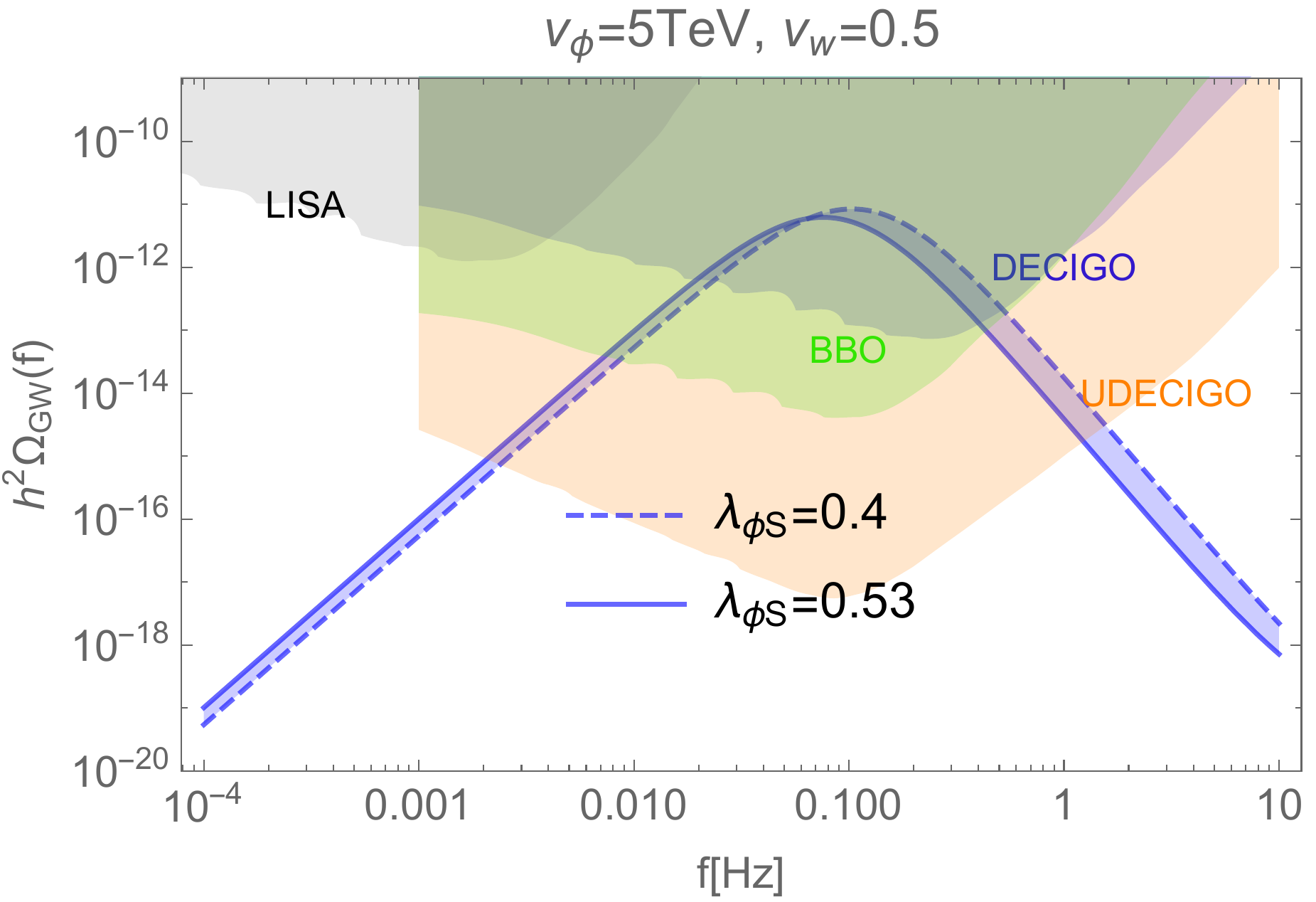}
\end{center}
\caption{
GW spectra for $v_\phi^{}=3~$TeV (upper left), $4~$TeV (upper right), and $5~$TeV (lower). For each $v_\phi$, the results of taking the minimum and maximum allowed values of $\lambda_{\phi S}$ are shown as dashed and solid blue lines, respectively. The colored regions show the sensitivity curves for various future detectors. 
}
\label{fig:GW}
\end{figure}
We now have all the necessary inputs to calculate the GW signals from the FOPT. 
In this paper, we rely on the numerical fitting functions presented in Ref.~\cite{Caprini:2015zlo}.  
Here, let us once again clarify the relevant parameters in our model. 
As discussed above, the strength parameters $(\alpha,\beta)$ are functions of $\lambda_{\phi S}^{}$ and $v_\phi^{}$, and they are constrained by the XENON1T experiment and the theoretical perturbativity bound.
In particular, only a finite range of $\lambda_{\phi S}^{}$ is allowed for a given fixed value of $v_\phi^{}$ as shown in Table~\ref{tab:benchmark}.  

\

In Fig.~\ref{fig:GW}, we show the GW energy spectra for $v_\phi^{}=3~$TeV (upper left), $4~$TeV (upper right), and $5~$TeV (lower). For each $v_\phi$, the results of taking the minimum and maximum allowed values of $\lambda_{\phi S}$ are shown as dashed and solid blue lines, respectively.
%
%
Here, the other parameters are chosen as 
\aln{v_\text{w}^{}
	&=	0.5,&
\epsilon
	&=	0.05,
}
where $\epsilon$ denotes the energy fraction of the turbulent bulk motion. 
The colored regions show the sensitivity curves of future detectors. 
One can see that the peak amplitudes can become as large as  ${\cal O}(10^{-12})$ with a peak frequency $10^{-2}\text{--}10^{ -1}$\,Hz, which can be tested by DECIGO and BBO.  
%

\section{Summary}\label{summary}
As a minimal model to explain the electroweak scale that is exponentially smaller than the Planck scale,
we have proposed the two-scalar extension of the SM 
with maximal multicriticality in the sense that all the superrenormalizable parameters of the scalar potential are fixed by the MPP.
As a first trial, we focused on the class of criticalities in which $S$ has the $\mathbb{Z}_2^{}$ symmetry $S\to-S$, the masses squared of $H$ and $S$ vanish, and the cubic couplings $\phi H^\dagger H$ and $\phi S^2$ vanish too. 
Then we can employ the classification of the maximally multicritical points of the two-scalar sector of $\phi$ and $S$ in Ref.~\cite{Kawai:2021lam}.
Among them, the so-called CP-1234 case is phenomenologically interesting, in which $S$ acts as dark matter, and a FOPT can occur around the electroweak scale in the early universe.

In this paper, we have studied the dark matter abundance and the constraints from its direct detection as well as from the LHC data and the theoretical perturbativity bound. 
As a result, we have found an allowed region with the masses of dark matter $m_S$ and the additional neutral scalar boson $m_{h_\phi}^{}$ to be 
$1.8\,\text{TeV}\lesssim m_S\lesssim 2.5\,\text{TeV}$ and $60\,\text{GeV}\lesssim m_S\lesssim 220\,\text{GeV}$, respectively. 
In addition,  the coupling constants are constrained to be $0.2\lesssim\lambda_{SH}^{}\lesssim 0.8$ and $0.1\lesssim \lambda_{\phi S}^{}\lesssim 1.1$.
Then, we have studied the FOPT and the resultant GW signals.  
One of the interesting features of the maximum criticality is the presence of the linear term in the $\phi$ potential, which ensures that the system does not have a long supercooling period.
We have calculated the phase-transition strength parameters and found $\alpha=10^{-2}\text{--}10^{-1}$ and $\beta/H=10^{2}\text{--}10^3$ within the allowed region of parameters summarized above.  
These moderate values of strength parameters imply that the GW signals are dominated by sound wave contributions.   
We have found that the peak amplitude can become as large as ${\cal O}(10^{-12})$ around the frequency $10^{-2}\text{--}10^{ -1}$\,Hz, which can be tested by future detectors such as DECIGO and BBO.

It would be interesting to study more general multicritical points of the two-scalar extension of the SM. As a different extension of the SM, the roles of $\phi$ and $S$ for the electroweak-scale generation can be played by a singlet complex scalar and another doublet Higgs, respectively. If this model has the Peccei-Quinn (PQ) symmetry~\cite{Peccei:1977hh,Wilczek:1977pj,Weinberg:1977ma,Kim:1979if,Shifman:1979if,Zhitnitsky:1980tq,Dine:1981rt,Kim:2008hd,diCortona:2015ldu}, it can be the DFSZ axion model. 
In general, PQ symmetry is difficult to impose because it is violated by anomalies at the ultraviolet cutoff scale. 
However, the MPP can have a possibility to realize such a pseudo-symmetry as one of its multicritical points. 
This will be left for future investigation. 

It would also be worth investigating the Higgs inflation in this model, and the effect of right-handed neutrinos, which makes the perturbativity bound milder, along the line of Ref.~\cite{Hamada:2021jls}.

\section*{Acknowledgements} 
We would like to thank Dr.~Toshinori Matsui for fruitful discussions.   
The work of YH is supported by JSPS Overseas Research Fellowships. 
The work of KO is partly supported by the JSPS Kakenhi Grant No.~19H01899. 
The work of KY is supported in part by the Grant-in-Aid for Early-Career Scientists, No.~19K14714. 
The work of KK is supported by Grant Korea NRF-2019R1C1CC1010050, 2019R1A6A1A10073437.  
H.K. thanks Professor Shin-Nan Yang and his family for their kind support through the Chin Yu chair professorship. 
The work of H.K. is also partially supported by the Japan Society of Promotion of Science, Grants (No. 20K03970 and 18H03708), by the Ministry of Science and Technology, R.O.C. (MOST 110-2811-M-002-500), and by National Taiwan University.

\appendix 
\def\thesection{Appendix \Alph{section}}

\bibliography{Bibliography}

\providecommand{\href}[2]{#2}\begingroup\raggedright\begin{thebibliography}{10}

\bibitem{LIGOScientific:2016aoc}
{\bfseries LIGO Scientific, Virgo} Collaboration, B.~P. Abbott {\em et~al.},
  ``{Observation of Gravitational Waves from a Binary Black Hole Merger},''
  \href{http://dx.doi.org/10.1103/PhysRevLett.116.061102}{{\em Phys. Rev.
  Lett.} {\bfseries 116} no.~6, (2016) 061102},
  \href{http://arxiv.org/abs/1602.03837}{{\ttfamily arXiv:1602.03837 [gr-qc]}}.

\bibitem{LIGOScientific:2017vwq}
{\bfseries LIGO Scientific, Virgo} Collaboration, B.~P. Abbott {\em et~al.},
  ``{GW170817: Observation of Gravitational Waves from a Binary Neutron Star
  Inspiral},'' \href{http://dx.doi.org/10.1103/PhysRevLett.119.161101}{{\em
  Phys. Rev. Lett.} {\bfseries 119} no.~16, (2017) 161101},
  \href{http://arxiv.org/abs/1710.05832}{{\ttfamily arXiv:1710.05832 [gr-qc]}}.

\bibitem{LIGOScientific:2018mvr}
{\bfseries LIGO Scientific, Virgo} Collaboration, B.~P. Abbott {\em et~al.},
  ``{GWTC-1: A Gravitational-Wave Transient Catalog of Compact Binary Mergers
  Observed by LIGO and Virgo during the First and Second Observing Runs},''
  \href{http://dx.doi.org/10.1103/PhysRevX.9.031040}{{\em Phys. Rev. X}
  {\bfseries 9} no.~3, (2019) 031040},
  \href{http://arxiv.org/abs/1811.12907}{{\ttfamily arXiv:1811.12907
  [astro-ph.HE]}}.

\bibitem{LIGOScientific:2021sio}
{\bfseries LIGO Scientific, VIRGO, KAGRA} Collaboration, R.~Abbott {\em
  et~al.}, ``{Tests of General Relativity with Gwtc-3},''
  \href{http://arxiv.org/abs/2112.06861}{{\ttfamily arXiv:2112.06861 [gr-qc]}}.

\bibitem{LIGOScientific:2021jlr}
{\bfseries LIGO Scientific, VIRGO, KAGRA} Collaboration, R.~Abbott {\em
  et~al.}, ``{All-Sky Search for Gravitational Wave Emission from Scalar Boson
  Clouds Around Spinning Black Holes in Ligo O3 Data},''
  \href{http://arxiv.org/abs/2111.15507}{{\ttfamily arXiv:2111.15507
  [astro-ph.HE]}}.

\bibitem{LIGOScientific:2017adf}
{\bfseries LIGO Scientific, Virgo, 1M2H, Dark Energy Camera GW-E, DES, DLT40,
  Las Cumbres Observatory, VINROUGE, MASTER} Collaboration, B.~P. Abbott {\em
  et~al.}, ``{A gravitational-wave standard siren measurement of the Hubble
  constant},'' \href{http://dx.doi.org/10.1038/nature24471}{{\em Nature}
  {\bfseries 551} no.~7678, (2017) 85--88},
  \href{http://arxiv.org/abs/1710.05835}{{\ttfamily arXiv:1710.05835
  [astro-ph.CO]}}.

\bibitem{NANOGrav:2020bcs}
{\bfseries NANOGrav} Collaboration, Z.~Arzoumanian {\em et~al.}, ``{The
  NANOGrav 12.5 yr Data Set: Search for an Isotropic Stochastic
  Gravitational-wave Background},''
  \href{http://dx.doi.org/10.3847/2041-8213/abd401}{{\em Astrophys. J. Lett.}
  {\bfseries 905} no.~2, (2020) L34},
  \href{http://arxiv.org/abs/2009.04496}{{\ttfamily arXiv:2009.04496
  [astro-ph.HE]}}.

\bibitem{Blasi:2020mfx}
S.~Blasi, V.~Brdar, and K.~Schmitz, ``{Has NANOGrav found first evidence for
  cosmic strings?},''
  \href{http://dx.doi.org/10.1103/PhysRevLett.126.041305}{{\em Phys. Rev.
  Lett.} {\bfseries 126} no.~4, (2021) 041305},
  \href{http://arxiv.org/abs/2009.06607}{{\ttfamily arXiv:2009.06607
  [astro-ph.CO]}}.

\bibitem{Vaskonen:2020lbd}
V.~Vaskonen and H.~Veerm\"ae, ``{Did NANOGrav see a signal from primordial
  black hole formation?},''
  \href{http://dx.doi.org/10.1103/PhysRevLett.126.051303}{{\em Phys. Rev.
  Lett.} {\bfseries 126} no.~5, (2021) 051303},
  \href{http://arxiv.org/abs/2009.07832}{{\ttfamily arXiv:2009.07832
  [astro-ph.CO]}}.

\bibitem{DeLuca:2020agl}
V.~De~Luca, G.~Franciolini, and A.~Riotto, ``{NANOGrav Data Hints at Primordial
  Black Holes as Dark Matter},''
  \href{http://dx.doi.org/10.1103/PhysRevLett.126.041303}{{\em Phys. Rev.
  Lett.} {\bfseries 126} no.~4, (2021) 041303},
  \href{http://arxiv.org/abs/2009.08268}{{\ttfamily arXiv:2009.08268
  [astro-ph.CO]}}.

\bibitem{Nakai:2020oit}
Y.~Nakai, M.~Suzuki, F.~Takahashi, and M.~Yamada, ``{Gravitational Waves and
  Dark Radiation from Dark Phase Transition: Connecting NANOGrav Pulsar Timing
  Data and Hubble Tension},''
  \href{http://dx.doi.org/10.1016/j.physletb.2021.136238}{{\em Phys. Lett. B}
  {\bfseries 816} (2021) 136238},
  \href{http://arxiv.org/abs/2009.09754}{{\ttfamily arXiv:2009.09754
  [astro-ph.CO]}}.

\bibitem{Addazi:2020zcj}
A.~Addazi, Y.-F. Cai, Q.~Gan, A.~Marciano, and K.~Zeng, ``{NANOGrav results and
  dark first order phase transitions},''
  \href{http://dx.doi.org/10.1007/s11433-021-1724-6}{{\em Sci. China Phys.
  Mech. Astron.} {\bfseries 64} no.~9, (2021) 290411},
  \href{http://arxiv.org/abs/2009.10327}{{\ttfamily arXiv:2009.10327
  [hep-ph]}}.

\bibitem{Grojean:2006bp}
C.~Grojean and G.~Servant, ``{Gravitational Waves from Phase Transitions at the
  Electroweak Scale and Beyond},''
  \href{http://dx.doi.org/10.1103/PhysRevD.75.043507}{{\em Phys. Rev. D}
  {\bfseries 75} (2007) 043507},
  \href{http://arxiv.org/abs/hep-ph/0607107}{{\ttfamily arXiv:hep-ph/0607107}}.

\bibitem{Leitao:2012tx}
L.~Leitao, A.~Megevand, and A.~D. Sanchez, ``{Gravitational waves from the
  electroweak phase transition},''
  \href{http://dx.doi.org/10.1088/1475-7516/2012/10/024}{{\em JCAP} {\bfseries
  10} (2012) 024}, \href{http://arxiv.org/abs/1205.3070}{{\ttfamily
  arXiv:1205.3070 [astro-ph.CO]}}.

\bibitem{Jinno:2015doa}
R.~Jinno, K.~Nakayama, and M.~Takimoto, ``{Gravitational waves from the first
  order phase transition of the Higgs field at high energy scales},''
  \href{http://dx.doi.org/10.1103/PhysRevD.93.045024}{{\em Phys. Rev. D}
  {\bfseries 93} no.~4, (2016) 045024},
  \href{http://arxiv.org/abs/1510.02697}{{\ttfamily arXiv:1510.02697
  [hep-ph]}}.

\bibitem{Hashino:2016rvx}
K.~Hashino, M.~Kakizaki, S.~Kanemura, and T.~Matsui, ``{Synergy between
  measurements of gravitational waves and the triple-Higgs coupling in probing
  the first-order electroweak phase transition},''
  \href{http://dx.doi.org/10.1103/PhysRevD.94.015005}{{\em Phys. Rev. D}
  {\bfseries 94} no.~1, (2016) 015005},
  \href{http://arxiv.org/abs/1604.02069}{{\ttfamily arXiv:1604.02069
  [hep-ph]}}.

\bibitem{Kajantie:1996mn}
K.~Kajantie, M.~Laine, K.~Rummukainen, and M.~E. Shaposhnikov, ``{Is there a~
  hot electroweak phase transition at $m_H \gtrsim m_W$?},''
  \href{http://dx.doi.org/10.1103/PhysRevLett.77.2887}{{\em Phys. Rev. Lett.}
  {\bfseries 77} (1996) 2887--2890},
  \href{http://arxiv.org/abs/hep-ph/9605288}{{\ttfamily arXiv:hep-ph/9605288}}.

\bibitem{Laine:1998jb}
M.~Laine and K.~Rummukainen, ``{What's new with the electroweak phase
  transition?},'' \href{http://dx.doi.org/10.1016/S0920-5632(99)85017-8}{{\em
  Nucl. Phys. B Proc. Suppl.} {\bfseries 73} (1999) 180--185},
  \href{http://arxiv.org/abs/hep-lat/9809045}{{\ttfamily
  arXiv:hep-lat/9809045}}.

\bibitem{Hamada:2020wjh}
Y.~Hamada, H.~Kawai, K.-y. Oda, and K.~Yagyu, ``{Dark Matter in Minimal
  Dimensional Transmutation with Multicritical-Point Principle},''
  \href{http://dx.doi.org/10.1007/JHEP01(2021)087}{{\em JHEP} {\bfseries 01}
  (2021) 087}, \href{http://arxiv.org/abs/2008.08700}{{\ttfamily
  arXiv:2008.08700 [hep-ph]}}.

\bibitem{Hamada:2021jls}
Y.~Hamada, H.~Kawai, K.~Kawana, K.-y. Oda, and K.~Yagyu, ``{Minimal Scenario of
  Criticality for Electroweak Scale, Neutrino Masses, Dark Matter, and
  Inflation},'' \href{http://dx.doi.org/10.1140/epjc/s10052-021-09735-z}{{\em
  Eur. Phys. J. C} {\bfseries 81} no.~11, (2021) 962},
  \href{http://arxiv.org/abs/2102.04617}{{\ttfamily arXiv:2102.04617
  [hep-ph]}}.

\bibitem{Coleman:1973jx}
S.~R. Coleman and E.~J. Weinberg, ``{Radiative Corrections as the Origin of
  Spontaneous Symmetry Breaking},''
  \href{http://dx.doi.org/10.1103/PhysRevD.7.1888}{{\em Phys. Rev. D}
  {\bfseries 7} (1973) 1888--1910}.

\bibitem{Kawai:2021lam}
H.~Kawai and K.~Kawana, ``{Multi-Critical Point Principle as the Origin of
  Classical Conformality and Its Generalizations},''
  \href{http://arxiv.org/abs/2107.10720}{{\ttfamily arXiv:2107.10720
  [hep-th]}}.

\bibitem{Haruna:2019zeu}
J.~Haruna and H.~Kawai, ``{Weak scale from Planck scale: Mass scale generation
  in a classically conformal two-scalar system},''
  \href{http://dx.doi.org/10.1093/ptep/ptz165}{{\em PTEP} {\bfseries 2020}
  no.~3, (2020) 033B01}, \href{http://arxiv.org/abs/1905.05656}{{\ttfamily
  arXiv:1905.05656 [hep-th]}}.

\bibitem{Bardeen:1995kv}
W.~A. Bardeen, ``{On naturalness in the standard model},'' in {\em {Ontake
  Summer Institute on Particle Physics}}.
\newblock 8, 1995.

\bibitem{Meissner:2006zh}
K.~A. Meissner and H.~Nicolai, ``{Conformal Symmetry and the Standard Model},''
  \href{http://dx.doi.org/10.1016/j.physletb.2007.03.023}{{\em Phys.Lett.}
  {\bfseries B648} (2007) 312--317},
\href{http://arxiv.org/abs/hep-th/0612165}{{\ttfamily arXiv:hep-th/0612165
  [hep-th]}}.

\bibitem{Meissner:2008gj}
K.~A. Meissner and H.~Nicolai, ``{Neutrinos, Axions and Conformal Symmetry},''
  \href{http://dx.doi.org/10.1140/epjc/s10052-008-0760-x}{{\em Eur. Phys. J. C}
  {\bfseries 57} (2008) 493--498},
  \href{http://arxiv.org/abs/0803.2814}{{\ttfamily arXiv:0803.2814 [hep-th]}}.

\bibitem{Foot:2007iy}
R.~Foot, A.~Kobakhidze, K.~L. McDonald, and R.~R. Volkas, ``{A Solution to the
  hierarchy problem from an almost decoupled hidden sector within a classically
  scale invariant theory},''
  \href{http://dx.doi.org/10.1103/PhysRevD.77.035006}{{\em Phys.Rev.}
  {\bfseries D77} (2008) 035006},
\href{http://arxiv.org/abs/0709.2750}{{\ttfamily arXiv:0709.2750 [hep-ph]}}.

\bibitem{Iso:2009ss}
S.~Iso, N.~Okada, and Y.~Orikasa, ``{Classically conformal $B-L$ extended
  Standard Model},''
  \href{http://dx.doi.org/10.1016/j.physletb.2009.04.046}{{\em Phys.Lett.}
  {\bfseries B676} (2009) 81--87},
\href{http://arxiv.org/abs/0902.4050}{{\ttfamily arXiv:0902.4050 [hep-ph]}}.

\bibitem{Iso:2009nw}
S.~Iso, N.~Okada, and Y.~Orikasa, ``{The minimal $B-L$ model naturally realized
  at TeV scale},'' \href{http://dx.doi.org/10.1103/PhysRevD.80.115007}{{\em
  Phys.Rev.} {\bfseries D80} (2009) 115007},
\href{http://arxiv.org/abs/0909.0128}{{\ttfamily arXiv:0909.0128 [hep-ph]}}.

\bibitem{Hur:2011sv}
T.~Hur and P.~Ko, ``{Scale invariant extension of the standard model with
  strongly interacting hidden sector},''
  \href{http://dx.doi.org/10.1103/PhysRevLett.106.141802}{{\em Phys.Rev.Lett.}
  {\bfseries 106} (2011) 141802},
\href{http://arxiv.org/abs/1103.2571}{{\ttfamily arXiv:1103.2571 [hep-ph]}}.

\bibitem{Iso:2012jn}
S.~Iso and Y.~Orikasa, ``{TeV Scale B-L model with a flat Higgs potential at
  the Planck scale - in view of the hierarchy problem -},''
  \href{http://dx.doi.org/10.1093/ptep/pts099}{{\em PTEP} {\bfseries 2013}
  (2013) 023B08},
\href{http://arxiv.org/abs/1210.2848}{{\ttfamily arXiv:1210.2848 [hep-ph]}}.

\bibitem{Englert:2013gz}
C.~Englert, J.~Jaeckel, V.~Khoze, and M.~Spannowsky, ``{Emergence of the
  Electroweak Scale through the Higgs Portal},''
  \href{http://dx.doi.org/10.1007/JHEP04(2013)060}{{\em JHEP} {\bfseries 04}
  (2013) 060}, \href{http://arxiv.org/abs/1301.4224}{{\ttfamily arXiv:1301.4224
  [hep-ph]}}.

\bibitem{Hashimoto:2013hta}
M.~Hashimoto, S.~Iso, and Y.~Orikasa, ``{Radiative symmetry breaking at the
  Fermi scale and flat potential at the Planck scale},''
  \href{http://dx.doi.org/10.1103/PhysRevD.89.016019}{{\em Phys.Rev.}
  {\bfseries D89} (2014) 016019},
\href{http://arxiv.org/abs/1310.4304}{{\ttfamily arXiv:1310.4304 [hep-ph]}}.

\bibitem{Holthausen:2013ota}
M.~Holthausen, J.~Kubo, K.~S. Lim, and M.~Lindner, ``{Electroweak and Conformal
  Symmetry Breaking by a Strongly Coupled Hidden Sector},''
  \href{http://dx.doi.org/10.1007/JHEP12(2013)076}{{\em JHEP} {\bfseries 12}
  (2013) 076}, \href{http://arxiv.org/abs/1310.4423}{{\ttfamily arXiv:1310.4423
  [hep-ph]}}.

\bibitem{Hashimoto:2014ela}
M.~Hashimoto, S.~Iso, and Y.~Orikasa, ``{Radiative Symmetry Breaking from Flat
  Potential in Various U(1)' Models},''
  \href{http://dx.doi.org/10.1103/PhysRevD.89.056010}{{\em Phys. Rev. D}
  {\bfseries 89} no.~5, (2014) 056010},
  \href{http://arxiv.org/abs/1401.5944}{{\ttfamily arXiv:1401.5944 [hep-ph]}}.

\bibitem{Kubo:2014ova}
J.~Kubo, K.~S. Lim, and M.~Lindner, ``{Electroweak Symmetry Breaking via
  QCD},'' \href{http://dx.doi.org/10.1103/PhysRevLett.113.091604}{{\em
  Phys.Rev.Lett.} {\bfseries 113} (2014) 091604},
\href{http://arxiv.org/abs/1403.4262}{{\ttfamily arXiv:1403.4262 [hep-ph]}}.

\bibitem{Endo:2015ifa}
K.~Endo and Y.~Sumino, ``{A Scale-invariant Higgs Sector and Structure of the
  Vacuum},'' \href{http://dx.doi.org/10.1007/JHEP05(2015)030}{{\em JHEP}
  {\bfseries 05} (2015) 030}, \href{http://arxiv.org/abs/1503.02819}{{\ttfamily
  arXiv:1503.02819 [hep-ph]}}.

\bibitem{Kubo:2015cna}
J.~Kubo and M.~Yamada, ``{Genesis of electroweak and dark matter scales from a
  bilinear scalar condensate},''
  \href{http://dx.doi.org/10.1103/PhysRevD.93.075016}{{\em Phys. Rev. D}
  {\bfseries 93} no.~7, (2016) 075016},
  \href{http://arxiv.org/abs/1505.05971}{{\ttfamily arXiv:1505.05971
  [hep-ph]}}.

\bibitem{Jung:2019dog}
D.-W. Jung, J.~Lee, and S.-H. Nam, ``{Scalar dark matter in the conformally
  invariant extension of the standard model},''
  \href{http://dx.doi.org/10.1016/j.physletb.2019.134823}{{\em Phys. Lett. B}
  {\bfseries 797} (2019) 134823},
  \href{http://arxiv.org/abs/1904.10209}{{\ttfamily arXiv:1904.10209
  [hep-ph]}}.

\bibitem{Bennett:1993pj}
D.~Bennett and H.~B. Nielsen, ``{Predictions for nonAbelian fine structure
  constants from multicriticality},''
  \href{http://dx.doi.org/10.1142/S0217751X94002090}{{\em Int. J. Mod. Phys. A}
  {\bfseries 9} (1994) 5155--5200},
  \href{http://arxiv.org/abs/hep-ph/9311321}{{\ttfamily arXiv:hep-ph/9311321}}.

\bibitem{Froggatt:1995rt}
C.~Froggatt and H.~B. Nielsen, ``{Standard model criticality prediction: Top
  mass 173 +- 5-GeV and Higgs mass 135 +- 9-GeV},''
  \href{http://dx.doi.org/10.1016/0370-2693(95)01480-2}{{\em Phys. Lett. B}
  {\bfseries 368} (1996) 96--102},
  \href{http://arxiv.org/abs/hep-ph/9511371}{{\ttfamily arXiv:hep-ph/9511371}}.

\bibitem{Nielsen:2012pu}
H.~B. Nielsen, ``{Predicted the Higgs Mass},'' {\em Bled Workshops Phys.}
  {\bfseries 13} no.~2, (2012) 94--126,
  \href{http://arxiv.org/abs/1212.5716}{{\ttfamily arXiv:1212.5716 [hep-ph]}}.

\bibitem{Kawai:2011qb}
H.~Kawai and T.~Okada, ``{Solving the Naturalness Problem by Baby Universes in
  the Lorentzian Multiverse},''
  \href{http://dx.doi.org/10.1143/PTP.127.689}{{\em Prog. Theor. Phys.}
  {\bfseries 127} (2012) 689--721},
  \href{http://arxiv.org/abs/1110.2303}{{\ttfamily arXiv:1110.2303 [hep-th]}}.

\bibitem{Kawai:2013wwa}
H.~Kawai, ``{Low energy effective action of quantum gravity and the naturalness
  problem},'' \href{http://dx.doi.org/10.1142/S0217751X13400010}{{\em Int. J.
  Mod. Phys. A} {\bfseries 28} (2013) 1340001}.

\bibitem{Hamada:2014ofa}
Y.~Hamada, H.~Kawai, and K.~Kawana, ``{Evidence of the Big Fix},''
  \href{http://dx.doi.org/10.1142/S0217751X14500997}{{\em Int. J. Mod. Phys. A}
  {\bfseries 29} (2014) 1450099},
  \href{http://arxiv.org/abs/1405.1310}{{\ttfamily arXiv:1405.1310 [hep-ph]}}.

\bibitem{Hamada:2014xra}
Y.~Hamada, H.~Kawai, and K.~Kawana, ``{Weak Scale From the Maximum Entropy
  Principle},'' \href{http://dx.doi.org/10.1093/ptep/ptv011}{{\em PTEP}
  {\bfseries 2015} (2015) 033B06},
  \href{http://arxiv.org/abs/1409.6508}{{\ttfamily arXiv:1409.6508 [hep-ph]}}.

\bibitem{Hamada:2015dja}
Y.~Hamada, H.~Kawai, and K.~Kawana, ``{Natural solution to the naturalness
  problem: The universe does fine-tuning},''
  \href{http://dx.doi.org/10.1093/ptep/ptv168}{{\em PTEP} {\bfseries 2015}
  no.~12, (2015) 123B03}, \href{http://arxiv.org/abs/1509.05955}{{\ttfamily
  arXiv:1509.05955 [hep-th]}}.

\bibitem{Kannike:2020qtw}
K.~Kannike, N.~Koivunen, and M.~Raidal, ``{Principle of Multiple Point
  Criticality in Multi-Scalar Dark Matter Models},''
  \href{http://dx.doi.org/10.1016/j.nuclphysb.2021.115441}{{\em Nucl. Phys. B}
  {\bfseries 968} (2021) 115441},
  \href{http://arxiv.org/abs/2010.09718}{{\ttfamily arXiv:2010.09718
  [hep-ph]}}.

\bibitem{Racioppi:2021ynx}
A.~Racioppi, J.~Rajasalu, and K.~Selke, ``{Multiple point criticality principle
  and Coleman-Weinberg inflation},''
  \href{http://arxiv.org/abs/2109.03238}{{\ttfamily arXiv:2109.03238
  [astro-ph.CO]}}.

\bibitem{Aprile:2018dbl}
{\bfseries XENON} Collaboration, E.~Aprile {\em et~al.}, ``{Dark Matter Search
  Results from a One Ton-Year Exposure of XENON1T},''
  \href{http://dx.doi.org/10.1103/PhysRevLett.121.111302}{{\em Phys. Rev.
  Lett.} {\bfseries 121} no.~11, (2018) 111302},
  \href{http://arxiv.org/abs/1805.12562}{{\ttfamily arXiv:1805.12562
  [astro-ph.CO]}}.

\bibitem{Belanger:2018ccd}
G.~B\'elanger, F.~Boudjema, A.~Goudelis, A.~Pukhov, and B.~Zaldivar,
  ``{micrOMEGAs5.0 : Freeze-in},''
  \href{http://dx.doi.org/10.1016/j.cpc.2018.04.027}{{\em Comput. Phys.
  Commun.} {\bfseries 231} (2018) 173--186},
  \href{http://arxiv.org/abs/1801.03509}{{\ttfamily arXiv:1801.03509
  [hep-ph]}}.

\bibitem{Aghanim:2018eyx}
{\bfseries Planck} Collaboration, N.~Aghanim {\em et~al.}, ``{Planck 2018
  results. VI. Cosmological parameters},''
  \href{http://dx.doi.org/10.1051/0004-6361/201833910}{{\em Astron. Astrophys.}
  {\bfseries 641} (2020) A6}, \href{http://arxiv.org/abs/1807.06209}{{\ttfamily
  arXiv:1807.06209 [astro-ph.CO]}}.

\bibitem{ATLAS:2021vrm}
{\bfseries ATLAS} Collaboration, ``{Combined measurements of Higgs boson
  production and decay using up to $139$ fb$^{-1}$ of proton-proton collision
  data at $\sqrt{s}= 13$ TeV collected with the ATLAS experiment},''.

\bibitem{Curtin:2016urg}
D.~Curtin, P.~Meade, and H.~Ramani, ``{Thermal Resummation and Phase
  Transitions},'' \href{http://dx.doi.org/10.1140/epjc/s10052-018-6268-0}{{\em
  Eur. Phys. J. C} {\bfseries 78} no.~9, (2018) 787},
  \href{http://arxiv.org/abs/1612.00466}{{\ttfamily arXiv:1612.00466
  [hep-ph]}}.

\bibitem{Senaha:2020mop}
E.~Senaha, ``{Symmetry Restoration and Breaking at Finite Temperature: An
  Introductory Review},'' \href{http://dx.doi.org/10.3390/sym12050733}{{\em
  Symmetry} {\bfseries 12} no.~5, (2020) 733}.

\bibitem{Parwani:1991gq}
R.~R. Parwani, ``{Resummation in a hot scalar field theory},''
  \href{http://dx.doi.org/10.1103/PhysRevD.45.4695}{{\em Phys. Rev. D}
  {\bfseries 45} (1992) 4695},
  \href{http://arxiv.org/abs/hep-ph/9204216}{{\ttfamily arXiv:hep-ph/9204216}}.
  [Erratum: Phys.Rev.D 48, 5965 (1993)].

\bibitem{Kolb:1990vq}
E.~W. Kolb and M.~S. Turner,
  \href{http://dx.doi.org/10.1201/9780429492860}{{\em {The Early Universe}}},
  vol.~69.
\newblock 1990.

\bibitem{Wang:2020jrd}
X.~Wang, F.~P. Huang, and X.~Zhang, ``{Phase transition dynamics and
  gravitational wave spectra of strong first-order phase transition in
  supercooled universe},''
  \href{http://dx.doi.org/10.1088/1475-7516/2020/05/045}{{\em JCAP} {\bfseries
  05} (2020) 045}, \href{http://arxiv.org/abs/2003.08892}{{\ttfamily
  arXiv:2003.08892 [hep-ph]}}.

\bibitem{Gouttenoire:2021kjv}
Y.~Gouttenoire, R.~Jinno, and F.~Sala, ``{Friction Pressure on Relativistic
  Bubble Walls},'' \href{http://arxiv.org/abs/2112.07686}{{\ttfamily
  arXiv:2112.07686 [hep-ph]}}.

\bibitem{Caprini:2015zlo}
C.~Caprini {\em et~al.}, ``{Science with the space-based interferometer eLISA.
  II: Gravitational waves from cosmological phase transitions},''
  \href{http://dx.doi.org/10.1088/1475-7516/2016/04/001}{{\em JCAP} {\bfseries
  04} (2016) 001}, \href{http://arxiv.org/abs/1512.06239}{{\ttfamily
  arXiv:1512.06239 [astro-ph.CO]}}.

\bibitem{Caprini:2019egz}
C.~Caprini {\em et~al.}, ``{Detecting gravitational waves from cosmological
  phase transitions with LISA: an update},''
  \href{http://dx.doi.org/10.1088/1475-7516/2020/03/024}{{\em JCAP} {\bfseries
  03} (2020) 024}, \href{http://arxiv.org/abs/1910.13125}{{\ttfamily
  arXiv:1910.13125 [astro-ph.CO]}}.

\bibitem{Guada:2020xnz}
V.~Guada, M.~Nemev\v{s}ek, and M.~Pintar, ``{FindBounce: Package for
  multi-field bounce actions},''
  \href{http://dx.doi.org/10.1016/j.cpc.2020.107480}{{\em Comput. Phys.
  Commun.} {\bfseries 256} (2020) 107480},
  \href{http://arxiv.org/abs/2002.00881}{{\ttfamily arXiv:2002.00881
  [hep-ph]}}.

\bibitem{Wainwright:2011kj}
C.~L. Wainwright, ``{CosmoTransitions: Computing Cosmological Phase Transition
  Temperatures and Bubble Profiles with Multiple Fields},''
  \href{http://dx.doi.org/10.1016/j.cpc.2012.04.004}{{\em Comput. Phys.
  Commun.} {\bfseries 183} (2012) 2006--2013},
  \href{http://arxiv.org/abs/1109.4189}{{\ttfamily arXiv:1109.4189 [hep-ph]}}.

\bibitem{Jinno:2016knw}
R.~Jinno and M.~Takimoto, ``{Probing a classically conformal B-L model with
  gravitational waves},''
  \href{http://dx.doi.org/10.1103/PhysRevD.95.015020}{{\em Phys. Rev. D}
  {\bfseries 95} no.~1, (2017) 015020},
  \href{http://arxiv.org/abs/1604.05035}{{\ttfamily arXiv:1604.05035
  [hep-ph]}}.

\bibitem{Espinosa:2010hh}
J.~R. Espinosa, T.~Konstandin, J.~M. No, and G.~Servant, ``{Energy Budget of
  Cosmological First-order Phase Transitions},''
  \href{http://dx.doi.org/10.1088/1475-7516/2010/06/028}{{\em JCAP} {\bfseries
  06} (2010) 028}, \href{http://arxiv.org/abs/1004.4187}{{\ttfamily
  arXiv:1004.4187 [hep-ph]}}.

\bibitem{Peccei:1977hh}
R.~Peccei and H.~R. Quinn, ``{CP Conservation in the Presence of Instantons},''
  \href{http://dx.doi.org/10.1103/PhysRevLett.38.1440}{{\em Phys. Rev. Lett.}
  {\bfseries 38} (1977) 1440--1443}.

\bibitem{Wilczek:1977pj}
F.~Wilczek, ``{Problem of Strong $P$ and $T$ Invariance in the Presence of
  Instantons},'' \href{http://dx.doi.org/10.1103/PhysRevLett.40.279}{{\em Phys.
  Rev. Lett.} {\bfseries 40} (1978) 279--282}.

\bibitem{Weinberg:1977ma}
S.~Weinberg, ``{A New Light Boson?},''
  \href{http://dx.doi.org/10.1103/PhysRevLett.40.223}{{\em Phys. Rev. Lett.}
  {\bfseries 40} (1978) 223--226}.

\bibitem{Kim:1979if}
J.~E. Kim, ``{Weak Interaction Singlet and Strong CP Invariance},''
  \href{http://dx.doi.org/10.1103/PhysRevLett.43.103}{{\em Phys. Rev. Lett.}
  {\bfseries 43} (1979) 103}.

\bibitem{Shifman:1979if}
M.~A. Shifman, A.~Vainshtein, and V.~I. Zakharov, ``{Can Confinement Ensure
  Natural CP Invariance of Strong Interactions?},''
  \href{http://dx.doi.org/10.1016/0550-3213(80)90209-6}{{\em Nucl. Phys. B}
  {\bfseries 166} (1980) 493--506}.

\bibitem{Zhitnitsky:1980tq}
A.~R. Zhitnitsky, ``{On Possible Suppression of the Axion Hadron Interactions.
  (In Russian)},'' {\em Sov. J. Nucl. Phys.} {\bfseries 31} (1980) 260.

\bibitem{Dine:1981rt}
M.~Dine, W.~Fischler, and M.~Srednicki, ``{A Simple Solution to the Strong CP
  Problem with a Harmless Axion},''
  \href{http://dx.doi.org/10.1016/0370-2693(81)90590-6}{{\em Phys. Lett. B}
  {\bfseries 104} (1981) 199--202}.

\bibitem{Kim:2008hd}
J.~E. Kim and G.~Carosi, ``{Axions and the Strong CP Problem},''
  \href{http://dx.doi.org/10.1103/RevModPhys.82.557}{{\em Rev. Mod. Phys.}
  {\bfseries 82} (2010) 557--602},
  \href{http://arxiv.org/abs/0807.3125}{{\ttfamily arXiv:0807.3125 [hep-ph]}}.
  [Erratum: Rev.Mod.Phys. 91, 049902 (2019)].

\bibitem{diCortona:2015ldu}
G.~Grilli~di Cortona, E.~Hardy, J.~Pardo~Vega, and G.~Villadoro, ``{The QCD
  axion, precisely},'' \href{http://dx.doi.org/10.1007/JHEP01(2016)034}{{\em
  JHEP} {\bfseries 01} (2016) 034},
  \href{http://arxiv.org/abs/1511.02867}{{\ttfamily arXiv:1511.02867
  [hep-ph]}}.

\end{thebibliography}\endgroup
\bibliographystyle{utphys}

\end{document}